\documentclass[a4paper,fleqn,usenatbib]{mnras}

\usepackage{newtxtext,newtxmath}

\usepackage[T1]{fontenc}
\usepackage{ae,aecompl}

\usepackage{graphicx}	
\usepackage{amsmath}	
\usepackage{amssymb}	
\usepackage{threeparttable}
\usepackage{booktabs}
\usepackage{verbatim}
\usepackage[all]{hypcap}
\usepackage[usenames, dvipsnames]{xcolor}
\usepackage[utf8]{inputenc}

\def\eck#1{\left\lbrack #1 \right\rbrack}

\def\rund#1{\left( #1 \right)}

\def\wave#1{\left\lbrace #1 \right\rbrace}
\def\ave#1{\left\langle #1 \right\rangle}

\font \bolditalics = cmmib12
\def\bx#1{\leavevmode\thinspace\hbox{\vrule\vtop{\vbox{\hrule\kern1pt
        \hbox{\vphantom{\tt/}\thinspace{\bf#1}\thinspace}}
      \kern1pt\hrule}\vrule}\thinspace}

\def \vc #1{{\textfont1=\bolditalics \hbox{$\bf#1$}}}

\title[Galaxy bias in KiDS+GAMA]{Unveiling Galaxy Bias via the Halo Model, KiDS and GAMA}

\author[A. Dvornik et al.]{Andrej Dvornik,$^{1}$\thanks{E-mail: dvornik@strw.leidenuniv.nl}
Henk Hoekstra,$^{1}$
Konrad Kuijken,$^{1}$
Peter Schneider,$^{2}$ \newauthor
Alexandra Amon,$^{3}$
Reiko Nakajima,$^{2}$
Massimo Viola,$^{1}$
Ami Choi,$^{4}$\newauthor
Thomas Erben,$^{2}$
Daniel J. Farrow,$^{5}$ 
Catherine Heymans,$^{3}$\newauthor
Hendrik Hildebrandt,$^{2}$
Crist{\'o}bal Sif{\'o}n,$^{6}$
Lingyu Wang$^{7,8}$
\\
$^{1}$Leiden Observatory, Leiden University, Niels Bohrweg 2, 2333 CA Leiden, The Netherlands.\\
$^{2}$Argelander-Institut f{\"u}r Astronomie, Auf dem H{\"u}gel 71, 53121 Bonn, Germany.\\
$^{3}$SUPA, Institute for Astronomy, University of Edinburgh, Royal Observatory, Blackford Hill, Edinburgh, EH9 3HJ, UK.\\
$^{4}$Center for Cosmology and AstroParticle Physics, The Ohio State University, 191 West Woodruff Avenue, Columbus, OH 43210, USA.\\
$^{5}$Max-Planck-Institut f{\"u}r extraterrestrische Physik, Postfach 1312 Giessenbachstrasse, D-85741 Garching, Germany.\\
$^{6}$Department of Astrophysical Sciences, Peyton Hall, Princeton University, Princeton, NJ 08544, USA.\\
$^{7}$SRON Netherlands Institute for Space Research, Landleven 12, 9747 AD Groningen, The Netherlands.\\
$^{8}$Kapteyn Astronomical Institute, University of Groningen, Postbus 800, 9700 AV Groningen, The Netherlands.
}

\date{Accepted XXX. Received YYY; in original form ZZZ}

\pubyear{2018}

\begin{document}
\label{firstpage}
\pagerange{\pageref{firstpage}--\pageref{lastpage}}
\maketitle

\begin{abstract}
We measure the projected galaxy clustering and galaxy-galaxy lensing signals using the Galaxy And Mass Assembly (GAMA) survey and Kilo-Degree Survey (KiDS) to study galaxy bias. We use the concept of non-linear and stochastic galaxy biasing in the framework of halo occupation statistics to constrain the parameters of the halo occupation statistics and to unveil the origin of galaxy biasing. The bias function $\Gamma_{\text{gm}}(r_{\text{p}})$, where $r_{\text{p}}$ is the projected comoving separation, is evaluated using the analytical halo model from which the scale dependence of $\Gamma_{\text{gm}}(r_{\text{p}})$, and the origin of the non-linearity and stochasticity in halo occupation models can be inferred. Our observations unveil the physical reason for the non-linearity and stochasticity, further explored using hydrodynamical simulations, with the stochasticity mostly originating from the non-Poissonian behaviour of satellite galaxies in the dark matter haloes and their spatial distribution, which does not follow the spatial distribution of dark matter in the halo. The observed non-linearity is mostly due to the presence of the central galaxies, as was noted from previous theoretical work on the same topic. We also see that overall, more massive galaxies reveal a stronger scale dependence, and out to a larger radius. Our results show that a wealth of information about galaxy bias is hidden in halo occupation models. These models should therefore be used to determine the influence of galaxy bias in cosmological studies. 

\end{abstract}




\section{Introduction}
\label{sec:intro}

In the standard cold dark matter and cosmological constant-dominated ($\Lambda$CDM) cosmological framework, galaxies form and reside within dark matter haloes, which themselves form from the highest density peaks in the initial Gaussian random density field \citep[e.g.][and references therein]{Mo2010}. In this case one expects that the spatial distribution of galaxies traces the spatial distribution of the underlying dark matter. Galaxies are however, biased tracers of the underlying dark matter distribution, because of the complexity of their evolution and formation \citep{Davis1985, Dekel1987, Cacciato2012a}. The relation between the distribution of galaxies and the underlying dark matter distribution, usually referred as \emph{galaxy bias}, is thus important to understand in order to properly comprehend galaxy formation and interpret studies that use galaxies as tracers of the underlying dark matter, particularly for those trying to constrain cosmological parameters.

If such a relation can be described with a single number $b$, the galaxy bias is linear and deterministic. As galaxy formation is a complex process, it would be naive to assume that the relation between the dark matter density field and galaxies is a simple one, described only with a single number. Such a relation might be non-linear (the relation between a galaxy and matter density fields cannot be described with only a single number), scale dependent (the galaxy bias is different on the different scales studied) or stochastic (the biasing relation has an intrinsic scatter around the mean value). Numerous authors have presented various arguments for why simple linear and deterministic bias is highly questionable \citep{Kaiser1984, Davis1985, Dekel1999}. Moreover, cosmological simulations and semi-analytical models suggest that galaxy bias takes a more complicated, non-trivial form \citep{Wang2008, Zehavi2011}. 

Observationally, there have been many attempts to test if galaxy bias is linear and deterministic. There have been studies relying on clustering properties of different samples of galaxies \citep[e.g.][]{Wang2008, Zehavi2011}, studies measuring high-order correlation statistics and ones directly comparing observed galaxy distribution fluctuations with the matter distribution fluctuations measured in numerical simulations \citep[see][and references therein]{Cacciato2012a}. What is more, there have also been observations combining galaxy clustering with weak gravitational (galaxy-galaxy) lensing measurements \citep{Hoekstra2002, Simon2007, Jullo2012, Buddendiek2016}. The majority of the above observations have confirmed that galaxy bias is neither linear nor deterministic \citep{Cacciato2012a}.

Even though the observational results are in broad agreement with theoretical predictions, until recently there was no direct connection between measurements and model predictions, mostly because the standard formalism used to define and predict the non-linearity and stochasticity of galaxy bias is hard to interpret in the framework of galaxy formation models. \citet{Cacciato2012a} introduced a new approach that allows for intuitive interpretation of galaxy bias, that is directly linked to galaxy formation theory and various concepts therein. They reformulated the galaxy bias description (and the non-linearity and stochasticity of the relation between the galaxies and underlying dark matter distribution) presented by \citet{Dekel1999} using the formalism of halo occupation statistics. As galaxies are thought to live in dark matter haloes, halo occupation distributions (a prescription on  how galaxies populate dark matter haloes) are a natural way to describe the galaxy-dark matter connection, and consequently the nature of galaxy bias. Combining the halo occupation distributions with the halo model \citep{Seljak2000, Peacock2000, Cooray2002, Bosch2012, Mead2015, Wibking2017}, allows us to compare observations to predictions of those models, which has the potential to unveil the \emph{hidden factors} -- sources of deviations from the linear and deterministic biasing \citep{Cacciato2012a}. Recently \citet{Simon2017} also showed that the halo model contains important information about galaxy bias. In this paper, however, we demonstrate how the stochasticity of galaxy bias arises from two different sources; the first is the relation between dark matter haloes and the underlying dark matter field, and the second is the manner in which galaxies populate dark matter haloes. As in \citet{Cacciato2012a}, we will focus on the second source of stochasticity, which indeed can be addressed using a halo model combined with halo occupation distributions.

The aim of this paper is to measure the galaxy bias using state of the art galaxy surveys and constrain the nature of it using the halo occupation distribution formalism. The same formalism can provide us with insights on the sources of deviations from the linear and deterministic biasing and the results can be used in cosmological analyses using the combination of galaxy-galaxy lensing and galaxy clustering and those based on the cosmic shear measurements. In this paper we make use of the predictions of \citet{Cacciato2012a} and apply them to the measurements provided by the imaging Kilo-Degree Survey \citep[KiDS;][]{Kuijken2015, DeJong2015}, accompanied by the spectroscopic Galaxy And Mass Assembly (GAMA) survey \citep{Driver2011} in order to get a grasp of the features of galaxy bias that can be measured using a combination of galaxy clustering and galaxy-galaxy lensing measurements with high precision.

The outline of this paper is as follows. In Section \ref{sec:biasing} we recap the galaxy biasing formulation of \citet{Cacciato2012a}. In Section \ref{sec:halomodel_intro} we introduce the halo model, its ingredients and introduce the main observable, which is a combination of galaxy clustering and galaxy-galaxy lensing. In Section \ref{sec:data} we present the data and measurement methods used in our analysis. We present our galaxy biasing results in Section \ref{sec:results0}, together with comparison with simulations and discuss and conclude in Section \ref{sec:discussion}. In the Appendix, we detail the calculation of the analytical covariance matrix, and provide full pairwise posterior distributions of our derived halo model parameters. We also provide a detailed derivation of the connection between the galaxy-matter correlation and the galaxy-galaxy lensing signal, explaining the use of two different definitions of the critical surface mass density in the literature. We highlight the key differences between our expressions and those found in several recent papers.

Throughout the paper we use the following cosmological parameters entering in the calculation of the distances and in the halo model \citep{PlanckCollaboration2015}: $\Omega_{\text{m}} = 0.3089$, $\Omega_{\Lambda} = 0.6911$, $\sigma_{8} = 0.8159$, $n_{\text{s}} = 0.9667$ and $\Omega_{\text{b}} = 0.0486$. We also use $\overline{\rho}_{\text{m}}$ as the present day mean matter density of the Universe \mbox{($\overline{\rho}_{\text{m}} = \Omega_{\text{m}, 0} \, \rho_{\text{crit}}$}, where \mbox{$\rho_{\text{crit}} = 3H^{2}_{0}/(8\pi G)$} and the halo masses are defined as $M = 4\pi r_{\Delta}^3 \Delta \; \overline{\rho}_{\text{m}} / 3 $ enclosed by the radius $r_\Delta$ within which the mean density of the halo is $\Delta$ times $\overline{\rho}_{\text{m}}$, with $\Delta = 200$). All the measurements presented in the paper are in comoving units, and $\log$ and $\ln$ refer to the 10-based logarithm and the natural logarithm, respectively.


\section{Biasing}
\label{sec:biasing}

This paper closely follows the biasing formalism presented in \citet{Cacciato2012a}, and we refer the reader to that paper for a thorough treatment of the topic. Here we shortly recap the galaxy biasing formalism of \citet{Cacciato2012a} and correct a couple of typos that we discovered during the study of his work. In this formalism the mean biasing function $b(M)$ \citep[the equivalent of the mean biasing function $b(\delta_{m})$ as defined by][]{Dekel1999} is, using new variables: the number of galaxies in a dark matter halo, $N$, and the mass of a dark matter halo, $M$:
\begin{equation}\label{eq:linbias1}
b(M) \equiv {\overline{\rho}_{\text{m}} \over \overline{n}_{\text{g}}} \, {\langle N \vert M \rangle \over M} \,,
\end{equation}
where $\overline{n}_{\text{g}}$ is the average number density of galaxies and $\langle N \vert M \rangle$ is the mean of the halo occupation distribution for a halo of mass M, defined as:
\begin{equation}\label{NMaver}
\langle N \vert M \rangle = \sum_{N=0}^{\infty} N \, P(N \vert M)\,,
\end{equation}
where $P(N \vert M)$ is the halo occupation distribution. Note that in this case, the simple linear, deterministic biasing corresponds to:
\begin{equation}\label{lindetbias}
N = {\overline{n}_{\text{g}} \over \overline{\rho}_{\text{m}}} \, M\,,
\end{equation}
which gives the expected value of $b(M) = 1$. As $N$ is an integer and the quantities $\overline{\rho}_{\text{m}}$, $\overline{n}_{\text{g}}$ and $M$ are in general non-integer, it is clear that in this formulation the linear, deterministic bias is unphysical. We define the moments of the bias function $b(M)$ as
\begin{equation}\label{moments}
\hat{b} \equiv {\langle b(M) M^{2} \rangle \over \langle M^{2} \rangle}\,, 
\end{equation}
and
\begin{equation}
\tilde{b}^{2} \equiv {\langle b^{2}(M) M^{2} \rangle \over \langle M^{2} \rangle}\,.
\end{equation}
where $\langle ... \rangle$\footnote{\citet{Cacciato2012a} used $\sigma_{M}^{2} \equiv \langle M^{2} \rangle$ throughout the paper, and we decided to drop the $\sigma_{M}^{2}$ for cleaner and more consistent equations.} indicates an effective average (an integral over dark matter haloes) defined in the following form:
\begin{equation}\label{eq:mass_average}
\langle x \rangle \equiv \int_{0}^{\infty} x \, n(M) \, \mathrm{d} M \,,
\end{equation}
where $n(M)$ is the halo mass function and $x$ is a property of the halo or galaxy population. In the case of linear bias, $b(M)$ is a constant and hence $\tilde{b}/\hat{b} = 1$. The same ratio, $\tilde{b}/\hat{b}$, is the relevant measure of the non-linearity of the biasing relation \citep{Dekel1999}. Its deviation from unity is a sign of a non-linear galaxy bias. From equation \ref{eq:linbias1} we can see that linear bias corresponds to halo occupation statistics for which $\langle N \vert M \rangle \propto M$.

In the same manner \citet{Cacciato2012a} also define the random halo bias of a single halo of mass $M$, that contains $N$ galaxies, as: 
\begin{equation}\label{epsNdef}
\varepsilon_N \equiv N - \langle N \vert M \rangle\,,
\end{equation}
which, by definition, will have a zero mean when averaged over all dark matter haloes, i.e. $\langle \varepsilon_N \vert M\rangle = 0$. This can be used to define the halo stochasticity function:
\begin{equation}\label{bsfmod}
\sigma_{b}^{2}(M) \equiv \left({\overline{\rho}_{\text{m}} \over \overline{n}_{\text{g}}} \right)^{2} \, 
{\langle \varepsilon^{2}_N \vert M \rangle \over \langle M^{2} \rangle} \,,
\end{equation}
from which, after averaging over halo mass, one gets the stochasticity parameter:
\begin{equation}\label{stochpardef}
\sigma_{b}^{2} \equiv \left({\overline{\rho}_{\text{m}} \over \overline{n}_{\text{g}}} \right)^{2} \,
{\langle \varepsilon^{2}_N \rangle \over \langle M^{2} \rangle} \,.
\end{equation}
If the stochasticity parameter $\sigma_{b} = 0$, then the galaxy bias is deterministic. In addition to the two bias moments $\tilde{b}$ and $\hat{b}$, one can also define some other bias parameters, particularly the ratio of the variances $b_{\text{var}}^{2} \equiv \langle \delta^{2}_{\text{g}} \rangle / \langle
\delta^{2}_{\text{m}} \rangle$ \citep{Dekel1999, Cacciato2012a}. Using this definition and an HOD-based formulation, \citet{Cacciato2012a} show that:
\begin{equation}\label{bvardef}
b_{\text{var}}^{2} = \left({\overline{\rho}_{\text{m}} \over \overline{n}_{\text{g}}} \right)^{2} \,
{\langle N^{2} \rangle \over \langle M^{2} \rangle}\,,
\end{equation}
where the averages are again calculated according to equation (\ref{eq:mass_average}). As the bias parameter is sensitive to both non-linearity and stochasticity, the total variance of the bias $b_{\text{var}}^{2}$ can also be written as:
\begin{equation}\label{bvaralt}
b_{\text{var}}^{2} = \tilde{b}^{2} + \sigma_{\text{b}}^{2}\,.
\end{equation}
Combining equation (\ref{bvardef}) and (\ref{bvaralt}) we find a relation for $\langle N^{2} \rangle$
\begin{equation}\label{eq:n2}
\langle N^{2} \rangle = \left(\frac{\overline{n}_{\text{g}}} {\overline{\rho}_{\text{m}}} \right)^{2} \,
\left[\tilde{b}^{2} + \sigma_{\text{b}}^{2} \right]\langle M^{2} \rangle \,.
\end{equation}
We can compare this to the covariance, which is obtained directly from equations (\ref{eq:linbias1}) and (\ref{lindetbias}):
\begin{equation}\label{eq:nm}
\langle NM \rangle = \frac{\overline{n}_{\text{g}}} {\overline{\rho}_{\text{m}}} \, \hat{b}\, \langle M^{2} \rangle \,.
\end{equation}
From all the equations above, it also directly follows that one can define a linear correlation coefficient as: $r \equiv {\langle N M \rangle / [\langle N^{2} \rangle \, \langle M^{2} \rangle]}$, such that, combining equations (\ref{eq:n2}) and (\ref{eq:nm}), $\hat{b}$ can be written as: $\hat{b} = b_{\text{var}}r$. 

This enables us to consider some special cases. The discrete nature of galaxies does not allow us to have galaxy bias that is both linear and deterministic \citep{Cacciato2012a}. Despite that, halo occupation statistics do allow bias that is linear and stochastic where;
\begin{eqnarray}\label{linstoch}
\hat{b}  = \tilde{b} = b(M) = 1 & \;\;\;\;\;\; & b_{\text{var}} = (1+\sigma_{\text{b}}^{2})^{1/2} \nonumber\\
\sigma_{\text{b}} \ne 0 & \;\;\;\;\;\; & r = (1+\sigma_{\text{b}}^{2})^{-1/2} \,.
\end{eqnarray}
or non-linear and deterministic;
\begin{eqnarray}\label{nldet}
\hat{b} \ne \tilde{b} \ne 1 & \;\;\;\;\;\; & b_{\text{var}} = \tilde{b} \nonumber\\
\sigma_{\text{b}} = 0 & \;\;\;\;\;\; & r = \hat{b}/\tilde{b} \ne 1 \,.
\end{eqnarray}
%

\section{Halo model}
\label{sec:halomodel_intro}

To express the HOD, we use the halo model, a successful analytic framework used to describe the clustering of dark matter and its evolution in the Universe \citep{Seljak2000, Peacock2000, Cooray2002, Bosch2012, Mead2015}. The halo model provides 
an ideal framework to describe the statistical weak lensing signal around a selection of galaxies, their clustering and cosmic shear signal. The halo model is built upon the statistical description of the properties of dark matter haloes (namely the average density profile, large scale bias and abundance) as well as on the statistical description of the galaxies residing in them. The halo model allows us to unveil the \emph{hidden} sources of bias stochasticity \citep{Cacciato2012a}. 

\subsection{Halo model ingredients}
\label{sec:halomodel_ingredients}

We assume that dark matter haloes are spherically symmetric, on average, and have density profiles, $\rho(r \vert M) = M \, u_{\text{h}}(r \vert M)$, that depend only on their mass $M$, and $u_{\text{h}}(r \vert M)$ is the normalised density profile of a dark matter halo. Similarly, we assume that satellite galaxies in haloes of mass $M$ follow a  spherical number density distribution $n_{\text{s}}(r \vert M) = N_{\text{s}} \, u_{\text{s}}(r \vert M)$, where $u_{\text{s}}(r \vert M)$ is the normalised density profile of satellite galaxies. Central galaxies always have $r=0$. We assume that the density profile of dark matter haloes follows an NFW profile \citep{Navarro1997}. Since centrals and satellites are distributed differently, we write the galaxy-galaxy power spectrum as:
\begin{equation}\label{Pkgalsplit}
P_{\text{gg}}(k) = f^{2}_{\text{c}} P_{\text{cc}}(k) + 
2 f_{\text{c}} f_{\text{s}} P_{\text{cs}}(k) + f^{2}_{\text{s}} P_{\text{ss}}(k)\,,
\end{equation}
while the galaxy-dark matter cross power spectrum is given by:
\begin{equation}\label{Pkgaldmsplit}
P_{\text{gm}}(k) = f_{\text{c}} P_{\text{cm}}(k) + f_{\text{s}} P_{\text{sm}}(k)\,.
\end{equation}
Here $f_{\text{c}} = \overline{n}_{\text{c}}/\overline{n}_{\text{g}}$ and $f_{\text{s}} =
\overline{n}_{\text{s}}/\overline{n}_{\text{g}} = 1 - f_{\text{c}}$ are the central and satellite
fractions, respectively, and the average number densities $\overline{n}_{\text{g}}$,
$\overline{n}_{\text{c}}$ and $\overline{n}_{\text{s}}$ follow from:
\begin{equation}\label{averng}
\overline{n}_{\text{x}} = \int_{0}^{\infty} \langle N_{\text{x}} \vert M \rangle \, n(M) \, \text{d} M\,,
\end{equation}
where `x' stands for `g' (for galaxies), `c' (for centrals) or `s' (for satellites) and $n(M)$ is the halo mass function in the following form: 
\begin{equation}\label{eq:hmf}
n(M) = \frac{\overline{\rho}_{\text{m}}}{M^{2}} \nu f(\nu) \frac{\mathrm{d} \ln \nu}{\mathrm{d} \ln M}\,,
\end{equation}
with $\nu = \delta_{\text{c}} / \sigma(M)$, where $\delta_{\text{c}}$ is the critical overdensity for spherical collapse at redshift $z$, and $\sigma(M)$ is the mass variance. For $f(\nu)$ we use the form presented in \citet{Tinker2010}. In addition, it is common practice to split two-point statistics into a 1-halo term (both points are located in the same halo) and a 2-halo term (the two points are located in different haloes). The 1-halo terms are:
\begin{equation}
P^{\text{1h}}_{\text{cc}}(k) = {1 \over \overline{n}_{\text{c}}}\,,
\end{equation}
\begin{equation}
P^{\text{1h}}_{\text{ss}}(k) = \beta \int_{0}^{\infty} \mathcal{H}_{\text{s}}^{2}(k, M) \, n(M) \, \text{d} M\,,
\end{equation}
and all other terms are given by:
\begin{equation}
P^{\text{1h}}_{\text{xy}}(k) = \int_{0}^{\infty} \mathcal{H}_{\text{x}}(k, M) \, \mathcal{H}_{\text{y}}(k, M) \, n(M) \, \text{d} M\,.
\end{equation}
Here `x' and `y' are either `c' (for central), `s' (for satellite), or `m' (for matter), $\beta$ is a Poisson parameter which arises from considering a scatter in the number of satellite galaxies at fixed halo mass [in this case a free parameter -- we define the $\beta$ in detail using equations (\ref{stoch_sat}), (\ref{stoch_sat_poisson}) and (\ref{beta_def})] and we have defined
\begin{equation}
\mathcal{H}_{\text{m}}(k, M) = {M \over \overline{\rho}_{\text{m}}} \,  \tilde{u}_{\text{h}}(k \vert M)\,,
\end{equation}
\begin{equation}
\mathcal{H}_{\text{c}}(k, M) = {\langle N_{\text{c}} \vert M \rangle \over \overline{n}_{\text{c}}} \,,
\end{equation}
and
\begin{equation}
\mathcal{H}_{\text{s}}(k, M) = {\langle N_{\text{s}} \vert M \rangle \over \overline{n}_{\text{s}}} \,  \tilde{u}_{\text{s}}(k \vert M)\,,
\end{equation}
with $\tilde{u}_{\text{h}}(k \vert M)$ and $\tilde{u}_{\text{s}}(k \vert M)$ the Fourier transforms of the halo density profile and the satellite number density profile, respectively, both normalised to unity [$\tilde{u}(k \! \!= \!\! 0 \vert M) \!\! = \!\! 1 $]. The various 2-halo terms are given by:
\begin{align}\label{P2hcc}
{P^{\text{2h}}_{\text{x}\text{y}}(k) = P_{\text{lin}}(k) \,  \int_{0}^{\infty} \text{d} M_1 \, \mathcal{H}_{\text{x}}(k, M_1) \, b_{\text{h}}(M_1)\, n(M_1)} \nonumber \\\ 
\times \int_{0}^{\infty} \text{d} M_2 \, \mathcal{H}_{\text{y}}(k, M_2) \, b_{\text{h}}(M_2)\, n(M_2) \,,
\end{align}
where $P_{\text{lin}}(k)$ is the linear power spectrum, obtained using the \citet{Eisenstein1997} transfer function, and $b_\text{h}(M,z)$ is the halo bias function. Note that in this formalism, the matter-matter power spectrum simply reads:
\begin{equation}
P_{\text{mm}}(k) = P^{\text{1h}}_{\text{mm}}(k) + P^{\text{2h}}_{\text{mm}}(k) \, .
\end{equation}
The two-point correlation functions corresponding to these power-spectra are obtained by simple Fourier transformation:
\begin{equation}\label{xiFTfromPK}
\xi_{\text{xy}}(r) = {1 \over 2 \pi^{2}} \int_{0}^{\infty} P_{\text{xy}}(k) \, {\sin kr \over kr} \, k^{2} \, \text{d} k\,, 
\end{equation}

For the halo bias function, $b_{\text{h}}$, we use the fitting function from \citet{Tinker2010}, as it was obtained using the same numerical simulation from which the halo mass function was obtained. We have adopted the parametrization of the concentration-mass relation, given by \citet{Duffy2011}:
\begin{equation}
\label{eq:con_duffy}
c(M, z) = 10.14\; A_{\text{c}} \ \left[\frac{M}{(2\times 10^{12} M_{\odot}/h)}\right]^{- 0.081}\ (1+z)^{-1.01} \,,
\end{equation}
with a free normalisation $A_{\text{c}}$ that accounts for the theoretical uncertainties in the concentration-mass relation due to discrepancies in the numerical simulations (mostly resolution and cosmologies) from which this scaling is usually inferred \citep{Viola2015}. We allow for additional normalisation $A_{\text{s}}$ for satellites, such that
\begin{equation}
c_{\text{s}}(M, z) = A_{\text{s}}\, c(M, z)\,,
\end{equation} 
which governs how satellite galaxies are spatially distributed inside a dark matter halo and tests the assumption of satellite galaxies following the density distribution of the dark matter haloes. If $A_{\text{s}} \neq 1$, the galaxy bias will vary on small scales, as demonstrated by \citet{Cacciato2012a}.  

\subsection{Conditional stellar mass function}
\label{sec:halomodel_csmf}

In order to constrain the cause for the stochasticity, non-linearity and scale dependence of galaxy bias, we model the halo occupation statistics using the Conditional Stellar Mass Function \citep[CSMF, heavily motivated by][]{Yang2008a, Cacciato2009, Cacciato2012, Wang2013, VanUitert2016}. The CSMF, $\Phi(M_{\star} \vert M)$, specifies the average number of galaxies of stellar mass $M_{\star}$ that reside in a halo of mass $M$. In this formalism, the halo occupation statistics of central galaxies are defined via the function: 
\begin{equation}\label{CLFsplit}
\Phi(M_{\star} \vert M) = \Phi_{\text{c}}(M_{\star}  \vert M) + \Phi_{\text{s}}(M_{\star}  \vert M)\,.
\end{equation}
In particular, the CSMF of central galaxies is modelled as a log-normal,
\begin{equation}\label{phi_c}
\Phi_{\text{c}}(M_{\star}  \vert M) = {1 \over {\sqrt{2\pi} \, {\ln}(10)\, \sigma_{\text{c}} M_{\star} } 
}{\exp}\left[- { {\log(M_{\star} / M^{*}_{\text{c}} )^2 } \over 2\,\sigma_{\text{c}}^{2}} \right]\, \,,
\end{equation}
and the satellite term as a modified Schechter function,
\begin{equation}\label{phi_s}
\Phi_{\text{s}}(M_{\star}  \vert M) = { \phi^{*}_{\text{s}} \over M^{*}_{\text{s}}}\,
\left({M_{\star} \over M^{*}_{\text{s}}}\right)^{\alpha_{\text{s}}} \,
{\exp} \left[- \left ({M_{\star} \over M^{*}_{\text{s}}}\right )^2 \right] 
\,,
\end{equation}
where $\sigma_{\text{c}}$ is the scatter between stellar mass and halo mass and $\alpha_{\text{s}}$ governs the power law behaviour of satellite galaxies. Note that $M^{*}_{\text{c}}$, $\sigma_{\text{c}}$, $\phi^{*}_{\text{s}}$, $\alpha_{\text{s}}$ and
$M^{*}_{\text{s}}$ are, in principle, all functions of halo mass $M$. We assume that $\sigma_{\text{c}}$ and $\alpha_{\text{s}}$ are independent of the halo mass $M$. Inspired by \citet{Yang2008a}, we parametrise $M^{*}_{\text{c}}$, $M^{*}_{\text{s}}$ and $\phi^{*}_{\text{s}}$ as:
\begin{equation}\label{eq:CMF4}
M^{*}_{\text{c}}(M) = M_{0} \frac{(M/M_{1})^{\gamma_{1}}}{[1 + (M/M_{1})]^{\gamma_{1} - \gamma_{2}}}\,.
\end{equation}
\begin{equation}\label{eq:CMF5}
M_{\text{s}}^{*}(M) = 0.56\ M^{*}_{\text{c}}(M)\,,
\end{equation}
and
\begin{equation}\label{eq:CMF7}
\log[\phi_{\text{s}}^{*}(M)] = b_{0} + b_{1}(\log m_{12})\,,
\end{equation}
where $m_{12} = M/(10^{12}M_{\odot}h^{-1})$. The factor of $0.56$ is also inspired by \citet{Yang2008a} and further tests by \citet{VanUitert2016} showed that using this assumption does not significantly affect the results. We can see that {the stellar to halo mass relation for $M \ll M_{1}$ behaves as $M^{*}_{\text{c}} \propto M^{\gamma_{1}}$ and for $M \gg M_{1}$, $M^{*}_{\text{c}} \propto M^{\gamma_{2}}$, where $M_{1}$ is a characteristic mass scale and $M_{0}$ is a normalisation. Here $\gamma_{1}$, $\gamma_{2}$, $b_{0}$ and $b_{1}$ are all free parameters.

From the CSMF it is straightforward to compute the halo occupation numbers.  For example, the average number of galaxies with stellar masses in the range $M_{\star,1} \leq M_{\star}  \leq M_{\star,2}$ is thus given by:
\begin{equation}\label{HODfromCLF}
\langle N \vert M \rangle = \int_{M_{\star,1}}^{M_{\star,2}} \Phi(M_{\star}  \vert M) \, \text{d} M_{\star} \,.
\end{equation}
The distinction we have made here, by splitting galaxies into centrals or satellites, is required to illustrate the main source of non-linearity and scale dependence of galaxy bias (see results in Section \ref{sec:results0}). To explore this, we follow \citet{Cacciato2012a}, and define the random halo biases following similar procedure as in equation (\ref{epsNdef}):
\begin{equation}\label{randbias}
\varepsilon_{\text{c}} \equiv N_{\text{c}} - \langle N_{\text{c}} \vert M \rangle \quad \text{and} \quad \varepsilon_{\text{s}} \equiv N_{\text{s}} - \langle N_{\text{s}} \vert M \rangle\,,
\end{equation}
and the halo stochasticity functions for centrals and satellites are given by:
\begin{align}\label{stoch_cen}
\langle \varepsilon_{\text{c}}^{2} \vert M \rangle &=  \sum_{N_{\text{c}}=0}^{\infty} (N_{\text{c}} - \langle N_{\text{c}} \vert M \rangle)^{2}\, P(N_{\text{c}} \vert M) \nonumber \\\
&= \langle N_{\text{c}}^{2} \vert M \rangle - \langle N_{\text{c}} \vert M \rangle ^{2} \nonumber \\\
&= \langle N_{\text{c}} \vert M \rangle - \langle N_{\text{c}} \vert M \rangle ^{2} \,,
\end{align}
\begin{align}\label{stoch_sat}
\langle \varepsilon_{\text{s}}^{2} \vert M \rangle &=  \sum_{N_{\text{s}}=0}^{\infty} (N_{\text{s}} - \langle N_{\text{s}} \vert M \rangle)^{2}\, P(N_{\text{s}} \vert M) \nonumber \\\
&= \langle N_{\text{s}}^{2} \vert M \rangle - \langle N_{\text{s}} \vert M \rangle ^{2} \,,
\end{align}
where we have used the fact that $\langle N^{2}_{\text{c}} \vert M \rangle = \langle N_{\text{c}} \vert M \rangle$, which follows from the fact that $N_{\text{c}}$ is either zero or unity. We can see that central galaxies only contribute to the stochasticity if $\langle N_{\text{c}} \vert M \rangle < 1$. If $\langle N_{\text{c}} \vert M \rangle = 1$, then the HOD is deterministic and the stochasticity function $\langle \varepsilon_{\text{c}}^{2} \vert M \rangle = 0$. The CSMF, however, only specifies the first moment of the halo occupation distribution $P(N \vert M)$. For central galaxies this is not a problem, as $\langle N^{2}_{\text{c}} \vert M \rangle = \langle N_{\text{c}} \vert M \rangle$. For satellite galaxies, we use that 
\begin{equation}\label{stoch_sat_poisson}
\langle N_{\text{s}}^{2} \vert M \rangle = \beta(M) \langle N_{\text{s}} \vert M \rangle^{2} + 
\langle N_{\text{s}} \vert M \rangle \,,
\end{equation}
where $\beta(M)$ is the mass dependent Poisson parameter defined as: 
\begin{equation}\label{beta_def}
\beta(M) \equiv  {\langle N_{\text{s}}(N_{\text{s}} - 1) \vert M \rangle \over  \langle N_{\text{s}} \vert M \rangle^{2}} \,,
\end{equation}
which is unity if $P(N_{\text{s}} \vert M)$ is given by a Poisson distribution, larger than unity if the distribution is wider than a Poisson distribution (also called super-Poissonian distribution) or smaller than unity if the distribution is narrower than a Poisson distribution (also called sub-Poissonian distribution). If $\beta(M)$ is unity, then equation (\ref{stoch_sat}) takes a simple form $\langle \varepsilon_{\text{s}}^{2} \vert M \rangle = \langle N_{\text{s}} \vert M \rangle$.

In what follows we limit ourselves to cases in which $\beta(M)$ is independent of halo mass, i.e., $\beta(M) = \beta$, and we treat $\beta$ as a free parameter.

Even without an application to the data, we can already learn a lot about the nature of galaxy bias from combining the HOD and halo model approaches to galaxy biasing as described in Section \ref{sec:biasing}. As realistic HODs (as formulated above) differ strongly from the simple scaling $\langle N \vert M \rangle \propto M$ (equation \ref{lindetbias}, which gives the linear and deterministic galaxy bias), they will inherently predict a galaxy bias that is strongly non-linear. Moreover, this seems to be mostly the consequence of central galaxies for which $\langle N_{\text{c}} \vert M \rangle$ never follows a power law. Even the satellite occupation distribution $\langle N_{\text{s}} \vert M \rangle$ is never close to the power law form, due to a cut-off at the low mass end, as galaxies at certain stellar mass require a minimum mass for their host halo \citep[][see also Figure 2 therein]{Cacciato2012a}. Given the behaviour of the halo model and the HOD, the stochasticity of the galaxy bias could most strongly arise from the non-zero $\sigma_{\text{c}}$ in equation (\ref{phi_c}) and the possible non-Poissonian nature of the satellite galaxy distribution for less massive galaxies. For more massive galaxies the main source of stochasticity can be shot noise, which dominates the stochasticity function, $\sigma_{\text{b}}$ in equation (\ref{stochpardef}), when the number density of galaxies is small. We use those free parameters of the HOD in a fit to the data (see Section \ref{sec:data}), to constrain the cause for the stochasticity, non-linearity and scale dependence of galaxy bias.

\subsection{Projected functions}
\label{sec:halomodel_projected}

We can project the 3D bias functions as defined by \citet{Dekel1999, Cacciato2012a} into two-dimensional, projected analogues, which are more easily accessible observationally. We start by defining the matter-matter, galaxy-matter, and galaxy-galaxy projected surface densities as:
\begin{equation}\label{eq:xyprojdef}
\Sigma_{\text{xy}}(r_{\text{p}}) = 2  {\overline{\rho}_{\text{m}}} \, \int_{r_{\text{p}}}^{\infty} \xi_{\text{xy}}(r) \, { r\, \text{d} r \over \sqrt{r^2 - r_{\text{p}}^2}}\,,
\end{equation}
where `x' and `y' stand either for `g' or `m', and $r_{\text{p}}$ is the \emph{projected} separation, with the change from standard line-of-sight integration to the integration along the projected separation using an Abel tranformation.  We also define $\overline{\Sigma}_{\text{xy}}(< r_{\text{p}})$ as its average inside $r_{\text{p}}$:
\begin{equation}\label{averageSigma}
\overline{\Sigma}_{\text{xy}}(< r_{\text{p}})  = \frac{2}{r^2_{\text{p}}} \int_{0}^{r_{\text{p}}}\Sigma_{\text{xy}}(R') R'\, \mathrm{d}R'\,,
\end{equation}
which we use to define the excess surface densities (ESD)
\begin{equation}\label{ESDxy}
\Delta \Sigma_{\text{xy}}(r_{\text{p}}) = \overline{\Sigma}_{\text{xy}}(< r_{\text{p}}) - \Sigma_{\text{xy}}(r_{\text{p}})\,.
\end{equation}
We include the contribution of the stellar mass of galaxies to the lensing signal as a point mass approximation, which we can write as:
\begin{equation}
\label{eq:point_mass}
\Delta \Sigma^{\text{pm}}_{\text{gm}}(r_{\text{p}}) = \frac{M_{\star, \text{med}} }{\pi r_{\text{p}}^{2}} \,,
\end{equation}
where $M_{\star, \text{med}}$ is the median stellar mass of the selected galaxies obtained directly from the GAMA catalogue \citep[][see Section \ref{sec:lenses} and Table \ref{tab:sample_properties} for more details]{Taylor2011}. This stellar mass contribution is fixed by each of our samples. According to the checks performed, the inclusion of the stellar mass contribution to the lensing signal does not affect our conclusions.

The obtained projected surface densities can subsequently be used to define the projected, 2D analogues of the 3D bias functions \citep[$b_{\text{g}}^{\text{3D}}$, $\mathcal{R}_{\text{gm}}^{\text{3D}}$ and $\Gamma_{\text{gm}}^{\text{3D}}$, ][]{Dekel1999, Cacciato2012a} as:
\begin{equation}\label{eq:bgaldef2}
b_{\text{g}}(r_{\text{p}}) \equiv \sqrt{\frac{\Delta \Sigma_{\text{gg}}(r_{\text{p}})} {\Delta \Sigma_{\text{mm}}(r_{\text{p}})}} \,, 
\end{equation}
\begin{equation}\label{eq:CCCdef2}
{\mathcal{R}}_{\text{gm}}(r_{\text{p}}) \equiv \frac{\Delta \Sigma_{\text{gm}}(r_{\text{p}})} {\sqrt{\Delta \Sigma_{\text{gg}}(r_{\text{p}})\, \Delta \Sigma_{\text{mm}}(r_{\text{p}})}}\,,
\end{equation}
and
\begin{equation}\label{eq:Gammadef2}
\Gamma_{\text{gm}}(r_{\text{p}}) \equiv \frac{b_{\text{g}}(r_{\text{p}})} {{\mathcal{R}}_{\text{gm}}(r_{\text{p}})} = \frac{\Delta \Sigma_{\text{gg}}(r_{\text{p}})}{\Delta \Sigma_{\text{gm}}(r_{\text{p}})}\,. 
\end{equation}
In what follows we shall refer to these as the `projected bias functions'.

In the case of the galaxy-dark matter cross correlation, the excess surface density $\Delta \Sigma_{\text{gm}}(r_{\text{p}}) = \gamma_{\text{t}}(r_{\text{p}}) \ \Sigma_{\text{cr,com}}$, where $\gamma_{\text{t}}(r_{\text{p}})$ is the tangential shear, which can be measured observationally using galaxy-galaxy lensing, and $\Sigma_{\text{cr,com}}$ is the comoving critical surface mass density:\footnote{In \citet{Dvornik2017a}, the same definition was used in all the calculations and plots shown, but erroneously documented in the paper. The equations (6) and (9) of that paper should have the same form as equations (\ref{eq:SigmaCrit}) and (\ref{eq:crit_effective}), as discussed in Appendix \ref{sec:appendix2}.}:
\begin{equation}
\label{eq:SigmaCrit}
\Sigma_{\text{cr,com}}=\frac{c^2}{4\pi G (1+ z_{\rm l})^{2}} \frac{D(z_{\rm s})}{D(z_{\rm l})D(z_{\rm l},z_{\rm s})} \, ,
\end{equation}
where $D(z_{\rm l})$ is the angular diameter distance to the lens, $D(z_{\rm l}, z_{\rm s})$ is the angular diameter distance between the lens and the source and $D(z_{\rm s})$ is the angular diameter distance to the source. In Appendix \ref{sec:appendix2} we discuss the exact derivation of equation (\ref{eq:SigmaCrit}) and the implications of using different coordinates. In the case of the galaxy-galaxy autocorrelation we can write that
\begin{equation}\label{eq:projected_clustering}
\Delta \Sigma_{\text{gg}}(r_{\text{p}}) = \overline{\rho}_{\text{m}} \left[{2 \over r^{2}_{\text{p}}} \int_{0}^{r_{\text{p}}} w_{\text{p}}(R') \, R' \, \text{d} R' - w_{\text{p}}(r_{\text{p}})\right]\,,
\end{equation}
where $w_{\text{p}}(r_{\text{p}})$ is the projected galaxy correlation function, and $w_{\text{p}}(r_{\text{p}}) = \Sigma_{\text{gg}}(r_{\text{p}}) / \overline{\rho}_{\text{m}}$. It is immediately clear that $\Delta \Sigma_{\text{gg}}(r_{\text{p}})$ can be obtained from the projected correlation function $w_{\text{p}}(r_{\text{p}})$, which is routinely measured in large galaxy redshift surveys. 

In terms of the classical 3D bias functions $b_{\text{g}}^{\text{3D}}$, $\mathcal{R}_{\text{gm}}^{\text{3D}}$ and $\Gamma_{\text{gm}}^{\text{3D}}$ \citep{Cacciato2012a}, the galaxies can be unbiased with respect to the underlying dark matter distribution, if and only if the following conditions are true: they are not central galaxies, the occupation number of satellite galaxies obeys Poisson statistics ($\beta = 1$), the normalised number density profile of satellite galaxies is identical to the one of the dark matter, and the occupational number of satellites is directly proportional to halo mass as $\langle N_{\text{s}} \rangle = M {\overline{n}_{\text{s}} / \rho}$. When central galaxies are added to the above conditions, one expects a strong scale dependence on small scales, due to the fact that central galaxies are strongly biased with respect to dark matter haloes. In the case of a non-Poissonian satellite distribution, one still expects $b_{\text{g}}^{\text{3D}} = 1$ on large scales, but with a transition from 1 to $\beta$, roughly at the virial radius when moving towards the centre of the halo \citep[see also Figure 3 in][]{Cacciato2012a}. The same also holds for the case where the density profile of satellites follows that of dark matter \citep{Cacciato2012a}.

Given all these reasons, as already pointed out by \citet{Cacciato2012a}, one expects scale independence on large scales (at a value dependent on halo model ingredients), with the transition to scale dependence on small scales (due to the effects of central galaxies) around the 1-halo to 2-halo transition. The same holds for the projected bias functions ($b_{\text{g}}$, $\mathcal{R}_{\text{gm}}$ and $\Gamma_{\text{gm}}$), which also carry a wealth of information regarding the non-linearity and stochasticity of halo occupation statistics, and consequently, galaxy formation.

This is demonstrated in Figure \ref{fig:GAMMA_predictions} where we show the influence of different values of $\sigma_{\text{c}}$, $A_{\text{s}}$, $\alpha_{\text{s}}$ and $\beta$ on the bias function $\Gamma_{\text{gm}}$ as a function of stellar mass. From the predictions one can clearly see how the different halo model ingredients influence the bias function. The halo model predicts, as mentioned before, scale independence above $10\, \text{Mpc}/h$ and a significant scale dependence of galaxy bias on smaller scales, with the parameters $\alpha_{\text{s}}$, $A_{\text{s}}$ and $\beta$ having a significant influence at those scales. Any deviation from a pure Poissonian distribution of satellite galaxies will result in quite a significant feature at intermediate scales, therefore it would be a likely explanation for detected signs of stochasticity [as the deviation from unity will drive the stochasticity function $\sigma_{\text{b}}$ or alternatively $\varepsilon$ away from $0$, as can be seen from equations (\ref{randbias}) to (\ref{beta_def})]. In Figure \ref{fig:GAMMA_predictions} we also test the influence of having different $\Omega_{\text{m}}$ and $\sigma_{8}$ on the $\Gamma_{\text{gm}}$ bias function, as generally, any bias function is a strong function of those two parameters \citep{Dekel1999, Sheldon2004}. We test this by picking $4$ combinations of $\Omega_{\text{m}}$ and $\sigma_{8}$ drawn from the $1\sigma$ confidence contours of \citet{PlanckCollaboration2015} measurements of the two parameters. Given the uncertainties of those parameters and their negligible influence on the $\Gamma_{\text{gm}}$ bias function, the decision to fix the cosmology seems to be justified.

We would like to remind the reader, that our implementation of the halo model does not include the scale dependence of the halo bias and the halo-exclusion (mutual exclusiveness of the spatial distribution of the haloes). Not including those effects can introduce errors on the 1-halo to 2-halo transition region that can be as large as $50\%$ \citep{Cacciato2012a, Bosch2012}. However, the bias functions as defined using equations (\ref{eq:bgaldef2}) to (\ref{eq:Gammadef2}) are much more accurate and less susceptible to the uncertainties in the halo model, by being defined as ratios of the two-point correlation functions \citep{Cacciato2012a}.

Despite of this, we decided to estimate the halo model parameters and the nature of galaxy bias using the fit to the $\Delta \Sigma_{\text{gm}}(r_{\text{p}})$ and $w_{\text{p}}(r_{\text{p}})$ signals separately, rather than the ratio of the two (using the $\Gamma_{\text{gm}}$ bias function directly). This approach will still suffer from a possible bias due to the fact that we do not include the scale dependent halo bias or the halo-exclusion in our model. This choice is motivated purely by the fact that the covariance matrix that would account for the cross-correlations between the lensing and clustering measurements cannot be properly taken into account when fitting the $\Gamma_{\text{gm}}$ bias function directly. We investigate the possible bias in our results in Section \ref{sec:bias_investigation}.

\begin{figure*}
	\includegraphics[width=\textwidth]{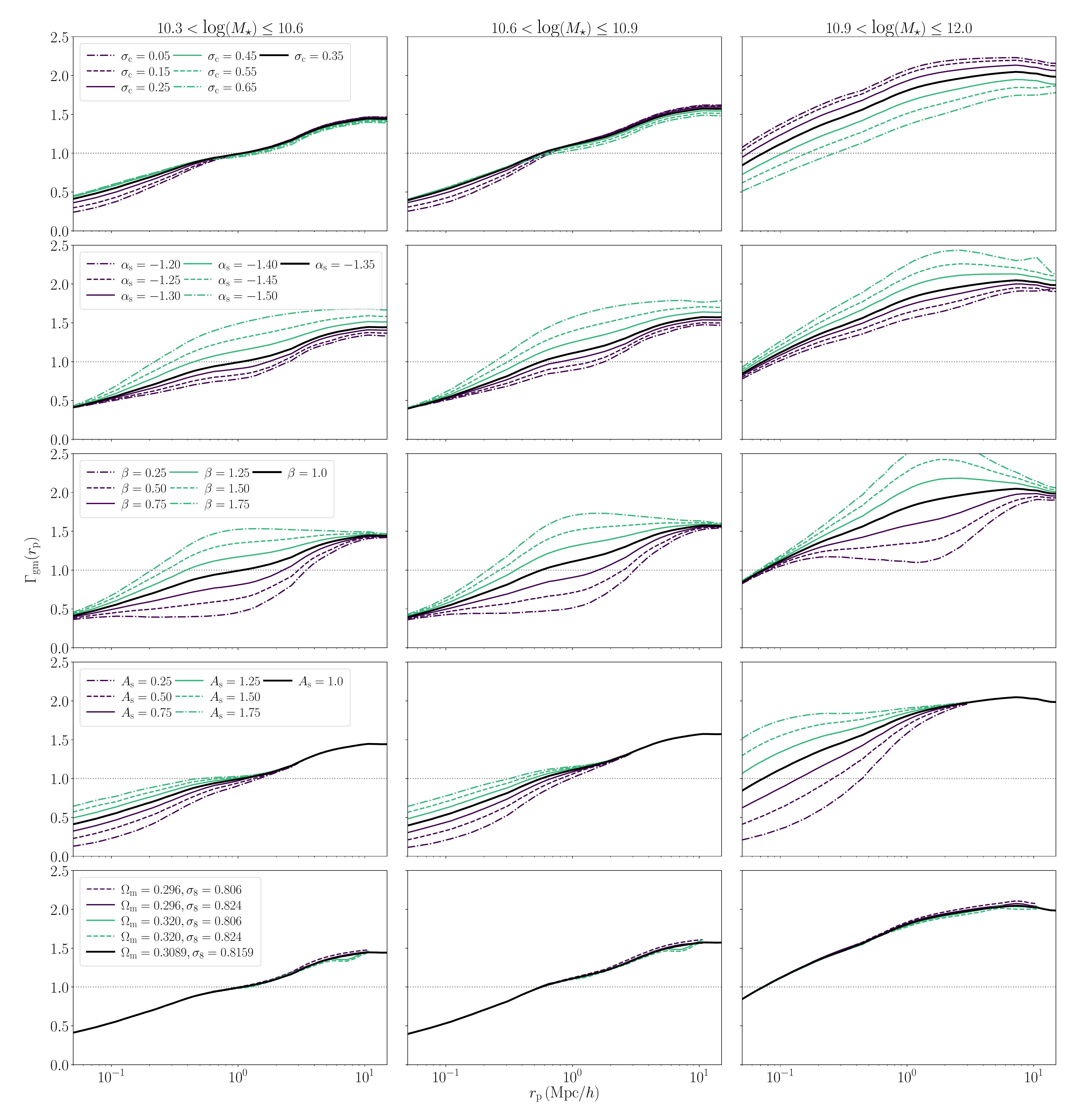}
 	\caption{Model predictions of scale dependence of the galaxy bias function $\Gamma_{\text{gm}}$ (equation \ref{eq:Gammadef2}) for three stellar mass bins (defined in Table \ref{tab:sample_properties}), with stellar masses given in units of $\left[\log(M_{\star} / [M_{\odot}/h^2])\right]$. With the black solid line we show our fiducial halo model \citep[with other parameters adapted from][]{Cacciato2012}, and the different green and violet lines show different values of $\sigma_{\text{c}}$, $\alpha_{\text{s}}$, $\beta$, $A_{\text{s}}$ and combinations of $\Omega_{\text{m}}$ and $\sigma_{8}$, row-wise, with values indicated in the legend. The full set of our fiducial parameters can be found in Table \ref{tab:results}.}
	\label{fig:GAMMA_predictions}
\end{figure*}


\section{Data and sample selection}
\label{sec:data}

\subsection{Lens galaxy selection}
\label{sec:lenses}

The foreground galaxies used in this lensing analysis are taken from the Galaxy And Mass Assembly (hereafter GAMA) survey \citep{Driver2011}. GAMA is a spectroscopic survey carried out on the Anglo-Australian Telescope with the AAOmega spectrograph. Specifically, we use the information of GAMA galaxies from three equatorial regions, G9, G12 and G15 from GAMA II \citep{Liske2015}. We do not use the G02 and G23 regions, because the first one does not overlap with KiDS and the second one uses a different target selection compared to the one used in the equatorial regions. These equatorial regions encompass \mbox{\textasciitilde{} 180 deg$^2$}, contain $180\,960$ galaxies (with $nQ \geq 3$, where the $nQ$ is a measure of redshift quality) and are highly complete down to a Petrosian $r$-band magnitude $r = 19.8$. For the weak lensing measurements, we use all the galaxies in the three equatorial regions as potential lenses.

To measure their average lensing and projected clustering signals, we group GAMA galaxies in stellar mass bins, following previous lensing measurements by \citet{VanUitert2016} and \citet{Velliscig2016}. The bin ranges were chosen this way to achieve a good signal-to-noise ratio in all bins and to measure the galaxy bias as a function of different stellar mass. The selection of galaxies can be seen in Figure \ref{fig:sample}, and the properties we use in the halo model are shown in Table \ref{tab:sample_properties}. Stellar masses are taken from version 19 of the stellar mass catalogue, an updated version of the catalogue created by \citet{Taylor2011}, who fitted \citet{Bruzual2003} synthetic stellar population SEDs to the broadband SDSS photometry assuming a \citet{Chabrier2003} IMF and a \citet{Calzetti2000} dust law. The stellar masses in \citet{Taylor2011} agree well with \mbox{MagPhys} derived estimates, as shown by \citet{Wright2017}. Despite the differences in the range of filters, star formation histories, obscuration laws, the two estimates agree within $0.2$ dex for 95 percent of the sample.

\begin{table}
	\caption{Overview of the median stellar masses of galaxies, median redshifts and number of galaxies/lenses in each selected bin, which are indicated in the second column. Stellar masses are given in units of $\left[\log(M_{\star} / [M_{\odot}/h^2])\right]$.}
	\label{tab:sample_properties}
	\centering
	\begin{tabular}{llllr} 
		\toprule
		Sample & Range & $M_{\star, \text{med}}$ & $z_{\text{med}}$ & \# of lenses\\
		\midrule
		Bin 1 & $(10.3, 10.6]$ & 10.46 & 0.244 & 26224\\
		Bin 2 & $(10.6, 10.9]$ & 10.74 & 0.284 & 20452\\
		Bin 3 & $(10.9, 12.0]$ & 11.13 & 0.318 & 10178\\
		\bottomrule
	\end{tabular}
\end{table}

\begin{figure}
	\includegraphics[width=\columnwidth]{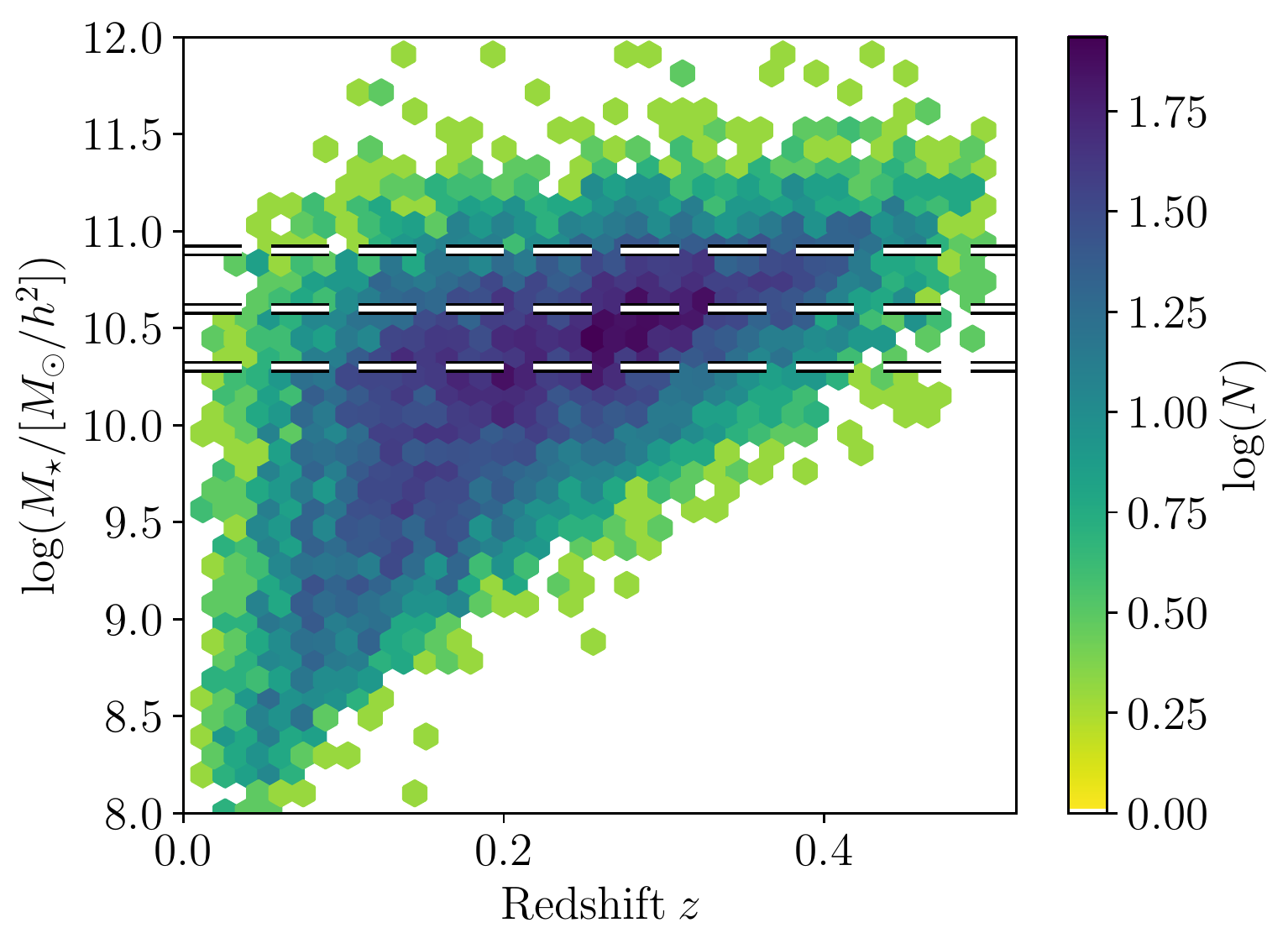}
 	\caption{Stellar mass versus redshift of galaxies in the GAMA survey that overlap with KiDS. The full sample is shown with hexagonal density plot and the dashed lines show the cuts for the three stellar mass bins used in our analysis.}
	\label{fig:sample}
\end{figure}
 
\subsection{Measurement of the $\Delta \Sigma_{\text{gm}}(r_{\text{p}})$ signal}
\label{sec:esd_measurement}

We use imaging data from $180$ deg$^2$ of KiDS \citep{Kuijken2015, DeJong2015} that overlaps with the GAMA survey \citep{Driver2011} to obtain shape measurements of background galaxies. KiDS is a four-band imaging survey conducted with the OmegaCAM CCD mosaic camera mounted at the Cassegrain focus of the VLT Survey Telescope (VST); the camera and telescope combination provide us with a fairly uniform point spread function across the field-of-view.

We use shape measurements based on the $r$-band images, which have an average seeing of $0.66$ arcsec. The image reduction, photometric redshift calibration and shape measurement analysis is described in detail in \citet{Hildebrandt2016}.

We measure galaxy shapes using calibrated \emph{lens}fit shape catalogs \citep{Miller2013} \citep[see also][where the calibration methodology is described]{Conti2016},
which provides galaxy ellipticities ($\mathrm{\epsilon_{1}}$, $\mathrm{\epsilon_{2}}$) with respect to an equatorial coordinate system. For each source-lens pair we compute the tangential $\epsilon_{\rm t}$ and cross component $\epsilon_{\times}$ of the source's ellipticity around the position of the lens:
\begin{equation}
\begin{bmatrix} \epsilon_{\rm t} \\
\epsilon_{\times}
\end{bmatrix}= \begin{bmatrix} -\cos(2\phi) & -\sin(2\phi) \\
\phantom{-}\sin(2\phi) & -\cos(2\phi) 
\end{bmatrix} \begin{bmatrix} \epsilon_{1} \\
\epsilon_{2}
\end{bmatrix}
,
\end{equation}
where $\mathrm{\phi}$ is the angle between the $x$-axis and the lens-source separation vector. 

The azimuthal average of the tangential ellipticity of a large number of galaxies in the same area of the sky is an unbiased estimate of the shear. On the other hand, the azimuthal average of the cross ellipticity over many sources is unaffected by gravitational lensing and should average to zero \citep{Schneider2003}. Therefore, the cross ellipticity is commonly used as an estimator of possible systematics in the measurements such as non-perfect PSF deconvolution, centroid bias and pixel level detector effects \citep{Mandelbaum2017}. Each lens-source pair is then assigned a weight
\begin{equation}
\label{eq:weights}
\widetilde{w}_{\mathrm{ls}}=w_{\mathrm{s}} \left(\widetilde \Sigma_{\mathrm{cr, ls}}^{-1}\right)^{2} \, ,
\end{equation}
which is the product of the \emph{lens}fit weight $w_{\text{s}}$ assigned to the given source ellipticity and the square of $\widetilde\Sigma_{\mathrm{cr, ls}}^{-1}$ -- the effective inverse critical surface mass density, which is a geometric term that downweights lens-source pairs that are close in redshift. We compute the effective inverse critical surface mass density for each lens using the spectroscopic redshift of the lens $z_{\mathrm{l}}$ and the full normalised redshift probability density of the sources, $n(z_{\mathrm{s}})$, calculated using the direct calibration method presented in \citet{Hildebrandt2016}.

The effective inverse critical surface density can be written as:
\begin{equation}
\label{eq:crit_effective}
\widetilde\Sigma_{\mathrm{cr, ls}}^{-1}=\frac{4\pi G}{c^2} (1+ z_{\rm l})^{2} D(z_{\mathrm{l}}) \int_{z_{\mathrm{l}}}^{\infty} \frac{D(z_{\mathrm{l}},z_{\mathrm{s}})}{D(z_{\mathrm{s}})}n(z_{\mathrm{s}})\, \mathrm{d}z_{\mathrm{s}} \, .
\end{equation}
The galaxy source sample is specific to each lens redshift with a minimum photometric redshift $z_{s} = z_{\mathrm{l}} +\delta_{z}$, with $\delta_{z} = 0.2$, where $\delta_{z}$ is an offset to mitigate the effects of contamination from the group galaxies \citep[for details see also the methods section and Appendix of][]{Dvornik2017a}. We determine the source redshift distribution $n(z_{\mathrm{s}})$ for each sample, by applying the sample photometric redshift selection to a spectroscopic catalogue that has been weighted to reproduce the correct galaxy colour-distributions in KiDS \citep[for details see][]{Hildebrandt2016}. Thus, the ESD can be directly computed in bins of projected distance $r_{\text{p}}$ to the lenses as:
\begin{equation}
\label{eq:ESDmeasured}
\Delta \Sigma_{\text{gm}} (r_{\text{p}}) = \left[ \frac{\sum_{\mathrm{ls}}\widetilde{w}_{\mathrm{ls}}\epsilon_{\mathrm{t, s}}\Sigma_{\mathrm{cr, ls}}^{\prime}}{\sum_{\mathrm{ls}}\widetilde{w}_{\mathrm{ls}}} \right] \frac{1}{1+\overline{m}} \, .
\end{equation}
where $\Sigma_{\mathrm{cr, ls}}^{\prime} \equiv 1/ \widetilde\Sigma_{\mathrm{cr, ls}}^{-1}$ and the sum is over all source-lens pairs in the distance bin, and
\begin{equation}
\overline{m} = \frac{\sum_{i}w_{i}^{\prime}m_{i}}{\sum_{i}w_{i}^{\prime}} \, ,
\end{equation}
is an average correction to the ESD profile that has to be applied to correct for the multiplicative bias $m$ in the \emph{lens}fit shear estimates. 
The sum goes over thin redshift slices for which $m$ is obtained using the method presented in \citet{Conti2016}, weighted 
by $w^{\prime} = w_{\rm s}D(z_{\rm l},z_{\rm s}) / D(z_{\rm s})$ for a given lens-source sample. The value of $\overline{m}$ is 
around $- 0.014$, independent of the scale at which it is computed. Furthermore, we subtract the signal around random points using the random catalogues from \citet{Farrow2015} \citep[for details see analysis in the Appendix of][]{Dvornik2017a}.

\subsection{Measurement of the $w_{\text{p}}(r_{\text{p}})$ profile}
\label{sec:wp_measurement}

We compute the three-dimensional autocorrelation function of our three lens samples using the \citet{Landy1993} estimator. For this we use the same random catalogue and procedure as described in \citet{Farrow2015}, applicable to the GAMA data. To minimise the effect of redshift-space distortions in our analysis, we project the three dimensional autocorrelation function along the line of sight:
\begin{equation}
\label{eq:projected_clustering}
w_{\text{p}}(r_{\text{p}}) = 2 \int_{0}^{\Pi_{\text{max}} = 100\, \text{Mpc}/h} \xi(r_{\text{p}}, \Pi)\, \text{d} \Pi \,.
\end{equation}
For practical reasons, the above integral is evaluated numerically. This calls for consideration of our integration limits, particularly the choice of $\Pi_{\text{max}}$. Theoretically one would like to integrate out to infinity in order to completely remove the effect of redshift space distortions and to encompass the full clustering signal on large scales. We settle for $\Pi_{\text{max}} = 100\, \text{Mpc}/h$, in order to project the correlation function on the separations we are interested in (with a maximum $r_{\text{p}} = 10\, \text{Mpc}/h$). We use the publicly available code \texttt{SWOT}\footnote{\url{http://jeancoupon.com/swot}} \citep{Coupon2012} to compute $\xi(r_{\text{p}}, \Pi)$ and $w_{\text{p}}(r_{\text{p}})$, and to get bootstrap estimates of the covariance matrix on small scales. The code was tested against results from \citet{Farrow2015} using the same sample of galaxies and updated random catalogues (internal version 0.3), reproducing the results in detail. Randoms generated by \citet{Farrow2015} contain around 750 times more galaxies than those in GAMA samples. Figure \ref{fig:randoms} shows the good agreement between the redshift distributions of the GAMA galaxies and the random catalogues for the three stellar mass bins.

The clustering signal $w_{\text{p}}(r_{\text{p}})$ as well as the lensing signal $\Delta \Sigma_{\text{gm}} (r_{\text{p}})$ are shown in Figure \ref{fig:ESD_bias}, in the right and left panel, respectively. They are shown together with MCMC best-fit profiles as described in Section \ref{sec:sampler}, using the halo model as described in Section \ref{sec:halomodel_intro}. The best-fit is a single model used for all stellar masses and not independent for the three bins we are using. In order to obtain the galaxy bias function $\Gamma_{\text{gm}}(r_{\text{p}})$ (equation \ref{eq:Gammadef2}) we project the clustering signal according to the equation (\ref{eq:projected_clustering}). The plot of this resulting function can be seen in Figure \ref{fig:GAMMA_full}. 

\begin{figure}
	\includegraphics[width=\columnwidth]{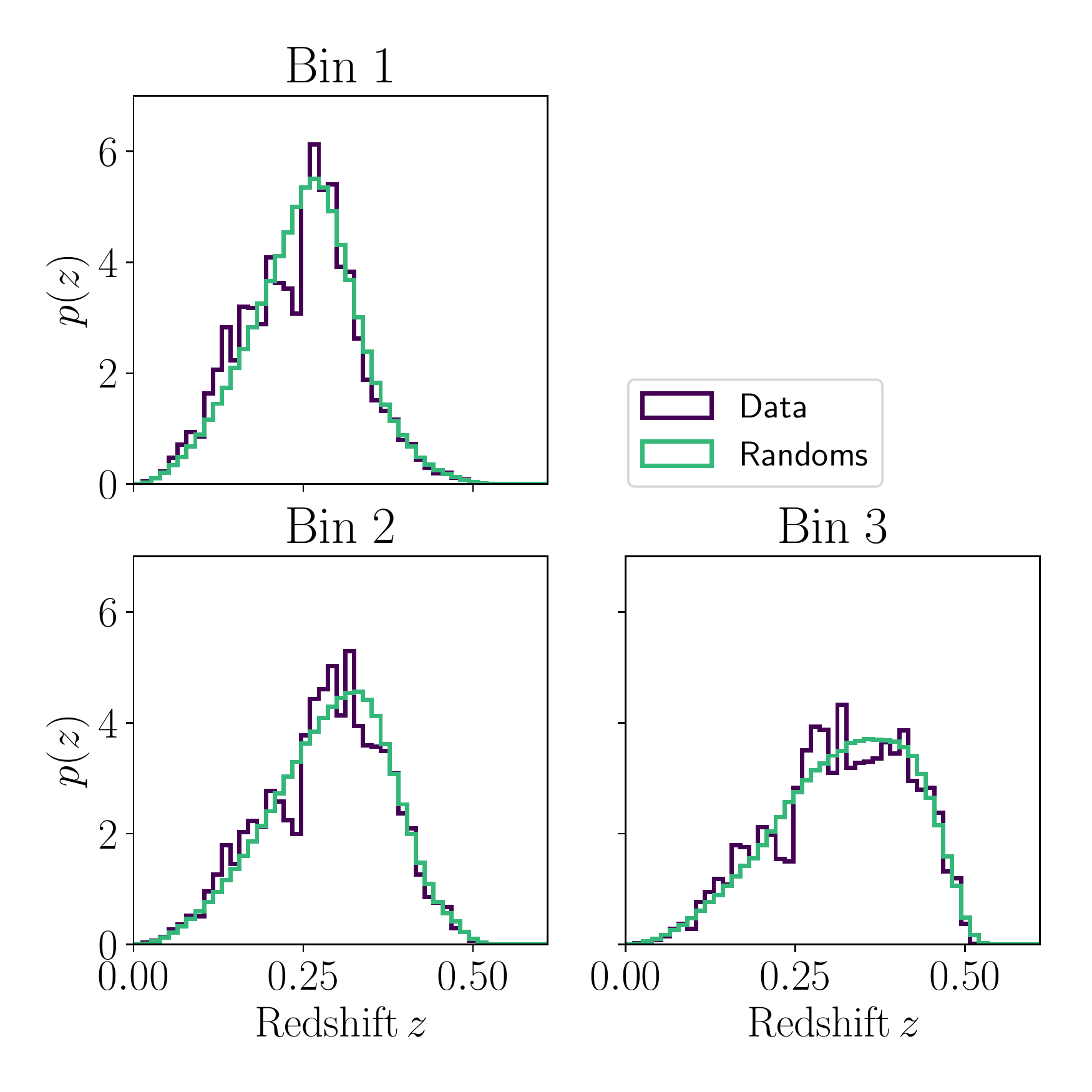}
 	\caption{A comparison between the redshift distribution of galaxies in the data and the matched galaxies in GAMA random catalogue \citep{Farrow2015} for our three stellar mass bins. We use the same set of randoms for both galaxy clustering and galaxy-galaxy lensing measurements.}
	\label{fig:randoms}
\end{figure}

\subsection{Covariance matrix estimation}
\label{sec:covariance}

Statistical error estimates on the lensing signal and projected galaxy clustering signal are obtained using an analytical covariance matrix. As shown in \citet{Dvornik2017a}, estimating the covariance matrix from data can become challenging given the small number of independent data patches in GAMA. This becomes even more challenging when one wants to include in the mixture the covariance for the projected galaxy clustering and all the possible cross terms between the two. The analytical covariance matrix we use is composed of three main parts: a Gaussian term, non-Gaussian term and the super-sample covariance (SSC) which accounts for all the modes outside of our KiDSxGAMA survey window. It is based on previous work by \citet{Takada2008}, \citet{Joachimi2008}, \citet{Pielorz2010},  \citet{Takada2013}, \citet{Li2014}, \citet{Marian2015}, \citet{Singh2016} and \citet{Krause2016}, and extended to support multiple lens bins and cross terms between lensing and projected galaxy clustering signals. The covariance matrix was tested against published results in these individual papers, as well as against real data estimates on small scales and mocks as used by \citet{VanUitert2017}. Further details and terms used can be found in Appendix \ref{sec:appendix}. We first evaluate our covariance matrix for a set of fiducial model parameters and use this in our MCMC fit and then take the best-fit values and re-evaluate the covariance matrix for the new best-fit halo model parameters. After carrying out the re-fitting procedure, we find out that the updated covariance matrix and halo model parameters do not affect the results of our fit, and thus the original estimate of the covariance matrix is appropriate to use throughout the analysis.

\subsection{Fitting procedure}
\label{sec:sampler}

The free parameters for our model are listed in Table \ref{tab:results}, together with their fiducial values. We use a Bayesian inference method in order to obtain full posterior  probabilities using a Monte Carlo Markov Chain (MCMC) technique; more specifically we use the \texttt{emcee}  Python package \citep{Foreman-Mackey2012}. The likelihood $\mathcal{L}$ is given by 
\begin{equation}\label{eq:likelihood} \mathcal{L} \propto \exp\left[- \frac{1}{2}(\boldsymbol{O}_{i}-\boldsymbol{M}_{i})^{T}\mathbfss{C}^{-1}_{ij}(\boldsymbol{O}_{j}-\boldsymbol{M}_{j})\right] \,, 
\end{equation} 
where $\boldsymbol{O}_{i}$ and $\boldsymbol{M}_{i}$ are the measurements and model predictions in radial bin $i$, and $\mathbfss{C}^{-1}_{ij}$ is the element of the inverse covariance matrix that accounts for the correlation between radial bins $i$ and $j$. In the fitting procedure we use the inverse covariance matrix as described in Section \ref{sec:covariance} and Appendix \ref{sec:appendix}. We use wide flat priors for all the parameters (given in Table \ref{tab:results}). The halo model (halo mass function and the power spectrum) is evaluated at the median redshift for each sample. 

We run the sampler using $120$ walkers, each with $12\,000$ steps (for a combined number of $14\,400\,000$ samples), out of which we discard the first $1000$ burn-in steps ($120\,000$ samples). The resulting MCMC chains are well converged according to the integrated autocorrelation time test.

\section{Results}
\label{sec:results0}

\subsection{KiDS and GAMA results}
\label{sec:results}

We fit the halo model as described in Section \ref{sec:sampler} to the measured projected galaxy clustering signal $w_{\text{p}}(r_{\text{p}})$ and the galaxy-galaxy lensing signal $\Delta \Sigma_{\text{gm}} (r_{\text{p}})$, using the covariance matrix as described in Section \ref{sec:covariance}. The resulting best fits are presented in Figure \ref{fig:ESD_bias} (together with the measurements and their respective $1\sigma$ errors obtained by taking the square root of the diagonal elements of the analytical covariance matrix). The measured halo model parameters, together with the $1\sigma$ uncertainties are summarised in Table \ref{tab:results}. Their full posterior distributions are shown in Figure \ref{fig:triangle}. The fit of our halo model to both the galaxy-galaxy lensing signal and projected galaxy clustering signal, using the full covariance matrix accounting for all the possible cross-correlations, has a reduced $\chi^{2}_{\text{red}} (\equiv  \chi^{2} / \text{d.o.f.})$ equal to 1.15, which is an appropriate fit, given the 33 degrees of freedom (d.o.f.). We urge readers not to rely on the ``chi-by-eye'' in Figures \ref{fig:ESD_bias} and \ref{fig:GAMMA_full} due to highly correlated data points (the correlations of which can be seen in Figure \ref{fig:cov_analytical_full}) and the joint fit of the halo model to the data.

\begin{table*}
	\caption{Summary of the lensing results obtained using MCMC halo model fit to the data. Here $M_{0}$ is the normalisation of the stellar to halo mass relation, $M_{1}$ is the characteristic mass scale of the same stellar to halo mass relation, $A_{\text{c}}$ is the normalisation of the concentration-mass relation, $\sigma_{\text{c}}$ is the scatter between the stellar and halo mass, $\gamma_{1}$ and $\gamma_{2}$ are the low and high-mass slopes of the stellar to halo mass relation, $A_{\text{s}}$ is the normalisation of the concentration-mass relation for satellite galaxies, $\alpha_{\text{s}}$, $b_{0}$ and $b_{1}$ govern the behaviour of the CSMF of satellite galaxies, and $\beta$ is the Poisson parameter. All parameters are defined in Section \ref{sec:halomodel_intro}, using equations (\ref{eq:con_duffy}) to (\ref{beta_def}).}
	\begin{threeparttable}
	\centering
	\label{tab:results}
	\begin{tabular}{lcccccc} 
		\toprule
		\quad & $\log(M_{0} / [M_{\odot}/h^{2}])$ & $\log(M_{1} / [M_{\odot}/h])$ & $A_{\text{c}}$ & $\sigma_{\text{c}}$ & $\gamma_{1}$ & $\gamma_{2}$\\
		\midrule
		\addlinespace
		Fiducial & $9.6$ & $11.25$ & $1.0$ & $0.35$ & $3.41$ & $0.99$\\
		\addlinespace
		Priors & $[7.0, 13.0]$\ & $[9.0, 14.0]$ & $[0.0, 5.0]$ & $[0.05, 2.0]$ & $[0.0, 10.0]$ & $[0.0, 10.0]$\\
		
		\addlinespace
		Posteriors & $8.75^{+1.62}_{-1.28}$ & $11.13^{+1.10}_{-1.11}$ & $1.33^{+0.20}_{-0.19}$ & $0.25^{+0.24}_{-0.18}$ & $2.16^{+4.43}_{-1.52}$ & $1.32^{+0.51}_{-0.34}$  \\
		\addlinespace
		\toprule
		\quad & $A_{\text{s}}$ & $\alpha_{\text{s}}$ & $b_{0}$ & $b_{1}$ & $\beta$ \\
		\midrule
		\addlinespace
		Fiducial & $1.0$ & $-1.34$ & $-1.15$ & $0.59$ & $1.0$\\
		\addlinespace
		Priors & $[0.0, 5.0]$ & $[-5.0, 5.0]$ & $[-5.0, 5.0]$ & $[-5.0, 5.0]$ & $[0.0, 2.0]$\\
		
		\addlinespace
		Posteriors & $0.24^{+0.30}_{-0.14}$ & $-1.36^{+0.19}_{-0.13}$ & $-0.71^{+0.34}_{-0.55}$ & $0.13^{+0.29}_{-0.30}$ & $1.67^{+0.15}_{-0.16}$ \\
		\addlinespace
		
		\bottomrule
	\end{tabular}
	\end{threeparttable}
\end{table*}

\begin{figure*}
	\includegraphics[width=\textwidth]{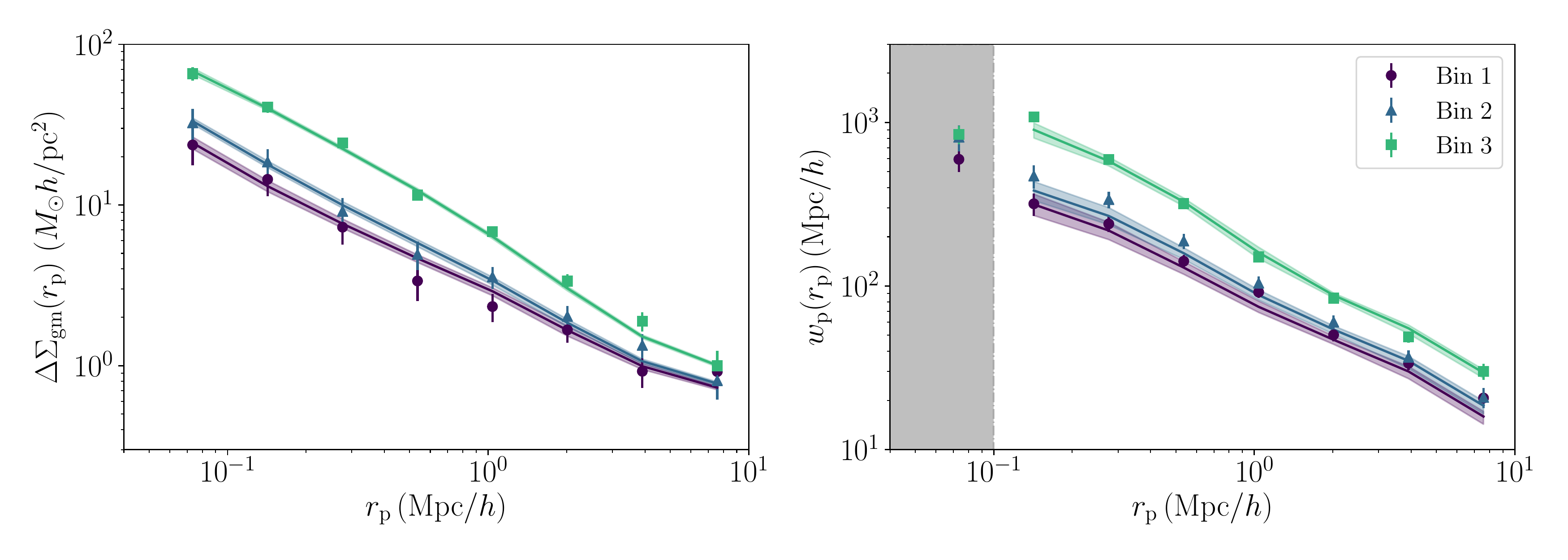}
 	\caption{The stacked ESD profile (\emph{left panel}) and projected galaxy clustering signal (\emph{right panel}) of the 3 stellar mass bins in the GAMA galaxy sample defined in Table \ref{tab:sample_properties}. The solid lines represent the best-fitting halo model as obtained using an MCMC fit, with the 68 percent confidence interval indicated with a shaded region. Using those two measurements we obtain the bias function $\Gamma_{\text{gm}}(r_{\text{p}})$. We do not use the measurements in the grey band in our fit, as the clustering measurements are affected by blending in this region. The best-fit halo model parameters are listed in Table \ref{tab:results}.}
	\label{fig:ESD_bias}
\end{figure*}

\begin{figure}
	\includegraphics[width=\columnwidth]{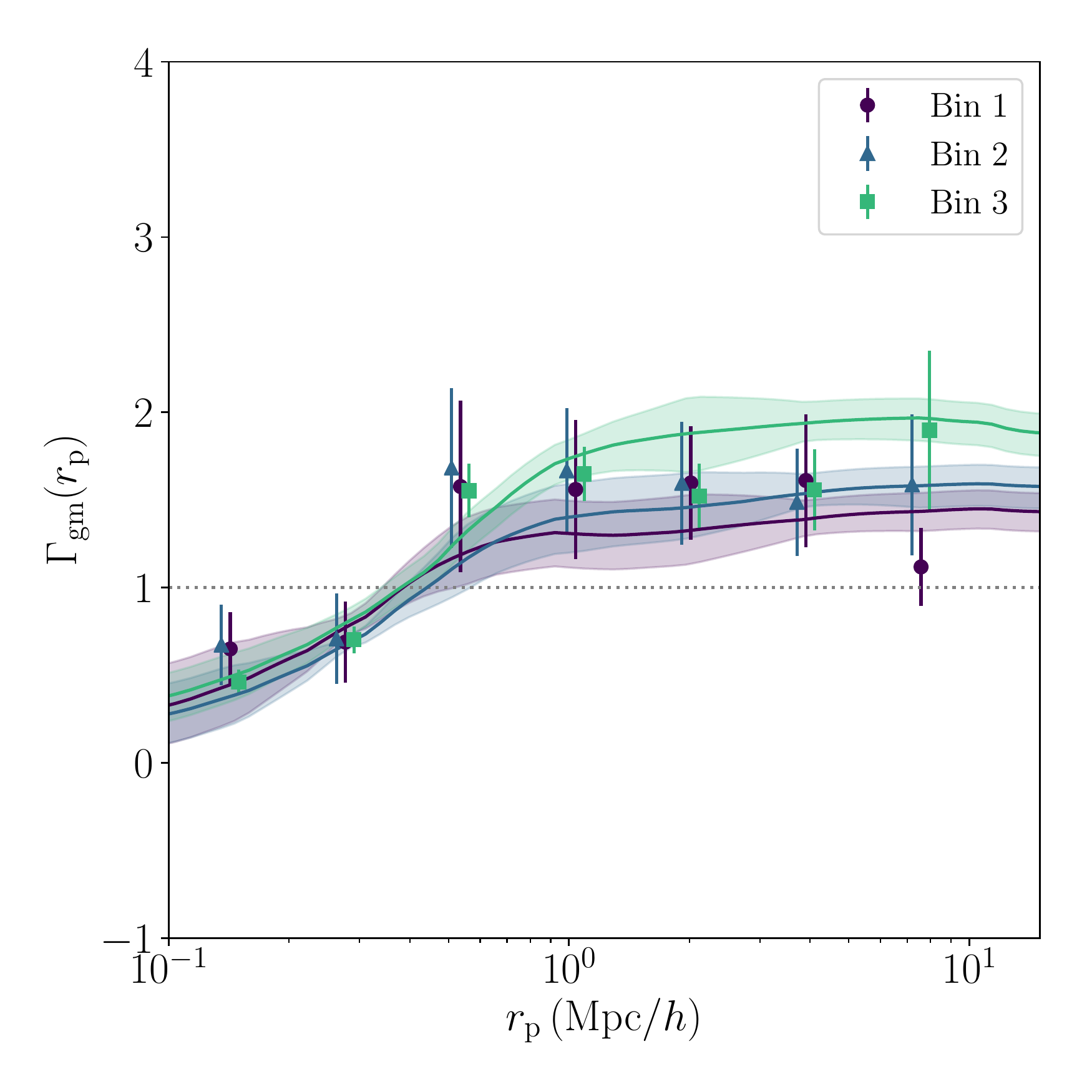}
 	\caption{The $\Gamma_{\text{gm}}(r_{\text{p}})$ bias function as measured using a combination of projected galaxy clustering and galaxy-galaxy lensing signals, shown for the 3 stellar mass bins as used throughout this paper. The solid lines represent the best-fitting halo model as obtained using an MCMC fit to the projected galaxy clustering and galaxy-galaxy lensing signal, combined to obtain $\Gamma_{\text{gm}}(r_{\text{p}})$, as described in Section \ref{sec:halomodel_intro}. The colour bands show the 68 percent confidence interval propagated from the best-fit model. Error bars on the data are obtained by propagating the appropriate sub-diagonals of the covariance matrix and thus do not show the correct correlations between the data points and also overestimate the sample variance and super-sample covariance contributions.}
	\label{fig:GAMMA_full}
\end{figure}

Due to the fact that we are only using samples with relatively high stellar masses, we are unable to sample the low-mass portion of the stellar mass function, evident in our inability to properly constrain the $\gamma_{1}$ parameter, which describes the behaviour of the stellar mass function at low halo mass. Mostly because of this, our results for the HOD parameters are different compared to those obtained by \citet{VanUitert2016}, who analysed the full GAMA sample. There is also a possible difference arising due to the available overlap of KiDS and GAMA surveys used in \citet{VanUitert2016} and our analysis, as \citet{VanUitert2016} used the lensing data from only $100$ deg$^{2}$ of the KiDS data, released before the shear catalogues used by \citet{Hildebrandt2016} and \citet{Dvornik2017a}, amongst others, became available. Our inferred HOD parameters are also in broad agreement with the ones obtained by \citet{Cacciato2013} for a sample of SDSS galaxies.

The main result of this work is the $\Gamma_{\text{gm}}(r_{\text{p}})$ bias function, presented in Figure \ref{fig:GAMMA_full}, together with the best fit MCMC result -- obtained by projecting the measured galaxy clustering result according to equation (\ref{eq:projected_clustering}) -- and combining with the galaxy-galaxy lensing result according to equation (\ref{eq:Gammadef2}). The obtained $\Gamma_{\text{gm}}(r_{\text{p}})$ bias function from the fit is scale dependent, showing a clear transition around $2\, \text{Mpc}/h$, in the 1-halo to 2-halo regime, where the function slowly transitions towards a constant value on even larger scales, beyond the range studied here \citep[as predicted in][]{Cacciato2012a}. Given the parameters obtained using the halo model fit to the data, the preferred value of $\beta$ is larger than unity with $\beta = 1.67^{+0.15}_{-0.16}$, which indicates that the satellite galaxies follow a super-Poissonian distribution inside their host dark matter haloes, and are thus responsible for the deviations from constant in our $\Gamma_{\text{gm}}(r_{\text{p}})$ bias function at intermediate scales. Following the formulation by \citet{Cacciato2012a}, this also means that the galaxy bias, as measured, is highly non-deterministic. As seen by the predictions shown in Figure \ref{fig:GAMMA_predictions}, the deviation of $\beta$ from unity alone is not sufficient to explain the full observed scale dependence of the $\Gamma_{\text{gm}}(r_{\text{p}})$ bias function. Given the best-fit parameter values using the MCMC fit of the halo model, the non-unity of the mass-concentration relation normalisation $A_{\text{s}}$ and other CSMF parameters (but most importantly the $\alpha_{\text{s}}$ parameter, which governs the power law behaviour of the satellite CSMF) are also responsible for the total contribution to the observed scale dependence, and thus the stochastic behaviour of the galaxy bias on all scales observed. 

\begin{figure}
	\includegraphics[width=\columnwidth]{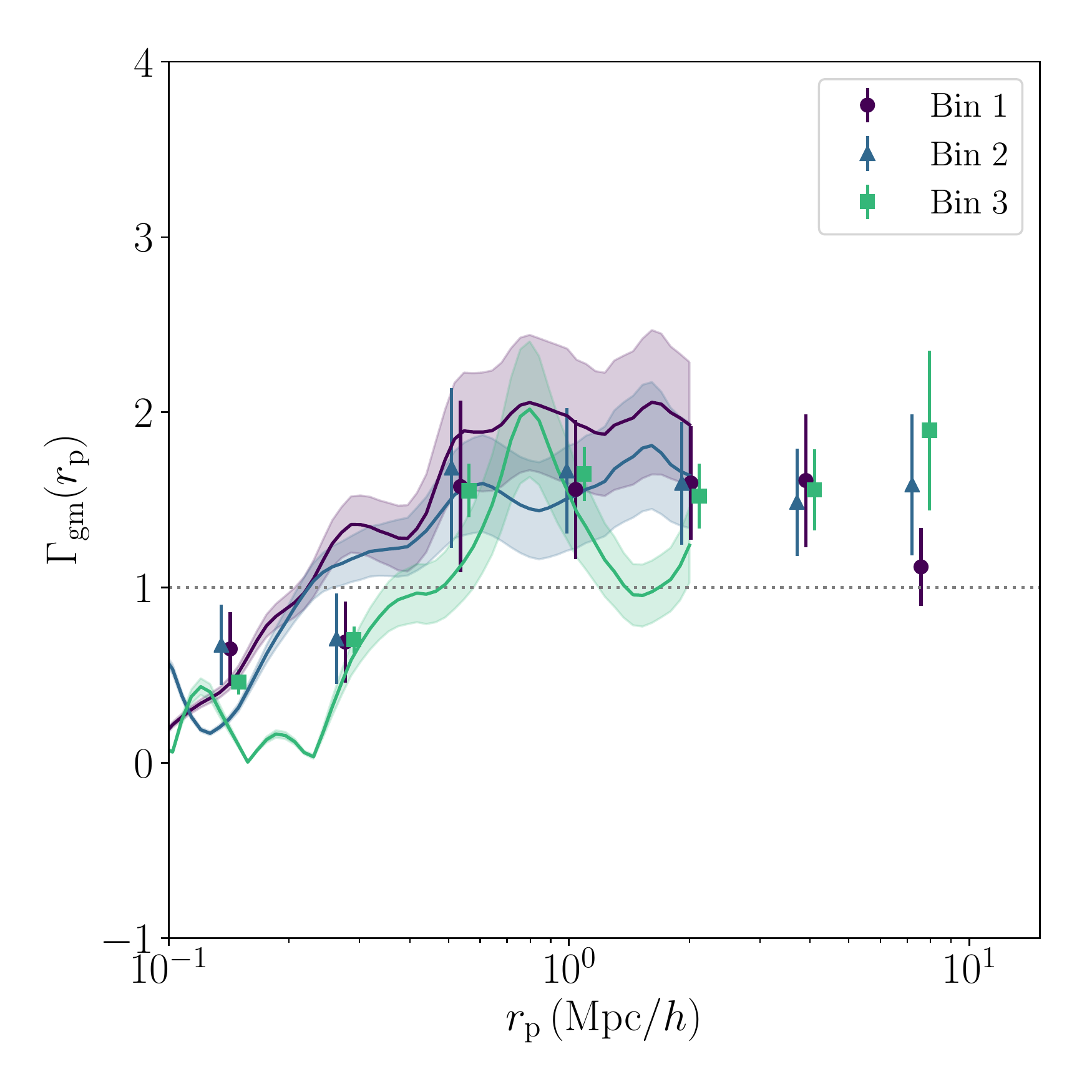}
 	\caption{The $\Gamma_{\text{gm}}(r_{\text{p}})$ bias function as measured using the combination of projected galaxy clustering and galaxy-galaxy lensing signals, shown for the 3 stellar mass bins as used throughout this paper. The solid lines represent the same measurement repeated on the EAGLE simulation, with the colour bands showing the $1\sigma$ errors. Note that those measurements are noisy due to the fact that the EAGLE simulation box is rather small, resulting in a relatively low number of galaxies in each bin (factor of around 26 lower, compared to the data). Due to the box size, we can also only show the measurement to about $2\, \text{Mpc}/h$.}
	\label{fig:GAMMA_eagle}
\end{figure}

\subsection{Investigation of the possible bias in the results}
\label{sec:bias_investigation}

Due to the fact that we have decided to fit the model to the $\Delta \Sigma_{\text{gm}}(r_{\text{p}})$ and $w_{\text{p}}(r_{\text{p}})$ signals, we investigate how this choice might have biased our results. To check this we repeat our analysis using the $\Gamma_{\text{gm}}(r_{\text{p}})$ bias function directly. As our data vector we take the ratio of the projected signals as shown in Figure \ref{fig:GAMMA_full} and we use the appropriately propagated sub-diagonals of the covariance matrix as a rough estimate of the total covariance matrix. Such a covariance matrix does not show the correct correlations between the data points (and the bins) and also overestimates the sample variance and super-sample covariance contributions. Nevertheless the ratio of the diagonals as an estimate of the errors is somewhat representative of the errors on the measured $\Gamma_{\text{gm}}(r_{\text{p}})$ bias function. The fit procedure (except for a different data vector, covariance and output of the model) follows the method presented in Section \ref{sec:sampler}. Using this, we obtain the best-fit values that are shown in Figure \ref{fig:triangle}, marked with blue points and lines, together with the full posterior distributions from the initial fit. The resulting fit has a $\chi^{2}_{\text{red}}$ equal to 1.29, with 9 degrees of freedom. As the results are consistent with the results that we obtain using a fit to the $\Delta \Sigma_{\text{gm}}(r_{\text{p}})$ and $w_{\text{p}}(r_{\text{p}})$ signals separately, it seems that, at least for this study, the halo model as described does not bias the overall conclusions of our analysis.

\subsection{Comparison with EAGLE simulation}
\label{sec:eagle}

In Figure \ref{fig:GAMMA_eagle} we compare our measurements of the GAMA and KiDS data to the same measurements made using the hydrodynamical EAGLE simulation \citep{Schaye2015, McAlpine2015}. EAGLE consists of state-of-the-art hydrodynamical simulations, including sub-grid interaction mechanisms between stellar and galactic energy sources. EAGLE is optimised such that the simulations reproduce a universe with the same stellar mass function as our own \citep{Schaye2015}. We follow the same procedure as with the data, by separately measuring the projected galaxy clustering signal and the galaxy-galaxy lensing signal and later combining the two accordingly. We measure the 3D galaxy clustering using the \citet{Landy1993} estimator, closely following the procedure outlined in \citet{Artale2017}. We adopt the same $\Pi_{\text{max}} = 34\, \text{Mpc}/h$ as used by \citet{Artale2017} in order to project the 3D galaxy clustering $\xi(r_{\text{p}}, \Pi)$ to $w_{\text{p}}(r_{\text{p}})$, which represents $\sim L/2$ of the EAGLE box \citep{Artale2017}; see also equation (\ref{eq:projected_clustering}). This limits the EAGLE measurements to a maximum scales of $r_{\text{p}} < 2\, \text{Mpc}/h$. As we do not require an accurate covariance matrix for the EAGLE results (we do not fit any model to it), we adopt a Jackknife covariance estimator using 8 equally sized sub-volumes. The measured EAGLE projected galaxy clustering signal is in good agreement with the GAMA measurements in detail, a result also found in \citet{Artale2017}.

To estimate the galaxy-galaxy lensing signal of galaxies in EAGLE, we use the excess surface density (i.e., lensing signal) of galaxies in EAGLE calculated by \citet{Velliscig2016}. We again select the galaxies in the three stellar mass bins, but in order to mimic the magnitude-limited sample we have adopted in our measurements of the galaxy-galaxy lensing signal on GAMA and KiDS, we have to weight our galaxies in the selection according to the satellite fraction as presented in \citet{Velliscig2016}.

Our two measurements (projected galaxy clustering and the galaxy-galaxy lensing) are then combined according to the definition of the $\Gamma_{\text{gm}}(r_{\text{p}})$ bias function, which is shown in Figure \ref{fig:GAMMA_eagle}. There we directly compare the bias function as measured in the KiDS and GAMA data to the one obtained from the EAGLE hydrodynamical simulation (shown with full lines). The results from EAGLE are noisy, due to the fact that one is limited by the number of galaxies present in EAGLE. 

Using the EAGLE simulations, we can directly access the properties of the satellite galaxies residing in the main halos present in the simulation. We select a narrow bin in halo masses of groups present in the simulation (between $12.0$ and $12.2$ in $\log(M/M_{\odot}$) and count the number of subhalos (galaxies). The resulting histogram, showing the relative abundance of satellite galaxies can be seen in Figure \ref{fig:eagle_poisson}. We also show the Poisson distribution with the same mean as the EAGLE data, as well as the Gaussian distribution with the same mean and standard deviation as the distribution of the satellite galaxies in our sample. It can be immediately seen that the distribution of satellite galaxies at a fixed halo mass does not follow a Poisson distribution, and it is significantly wider (thus indeed being super-Poissonian).

The comparison nevertheless shows that the galaxy bias is intrinsically scale dependent and the shape of it suggests that it can be attributed to the non-Poissonian behaviour of satellite galaxies (and to lesser extent also to the precise distribution of satellites in the dark matter halo, governed by $\alpha_{\text{s}}$ and $A_{\text{s}}$ in the halo model).

\begin{figure}
	\includegraphics[width=\columnwidth]{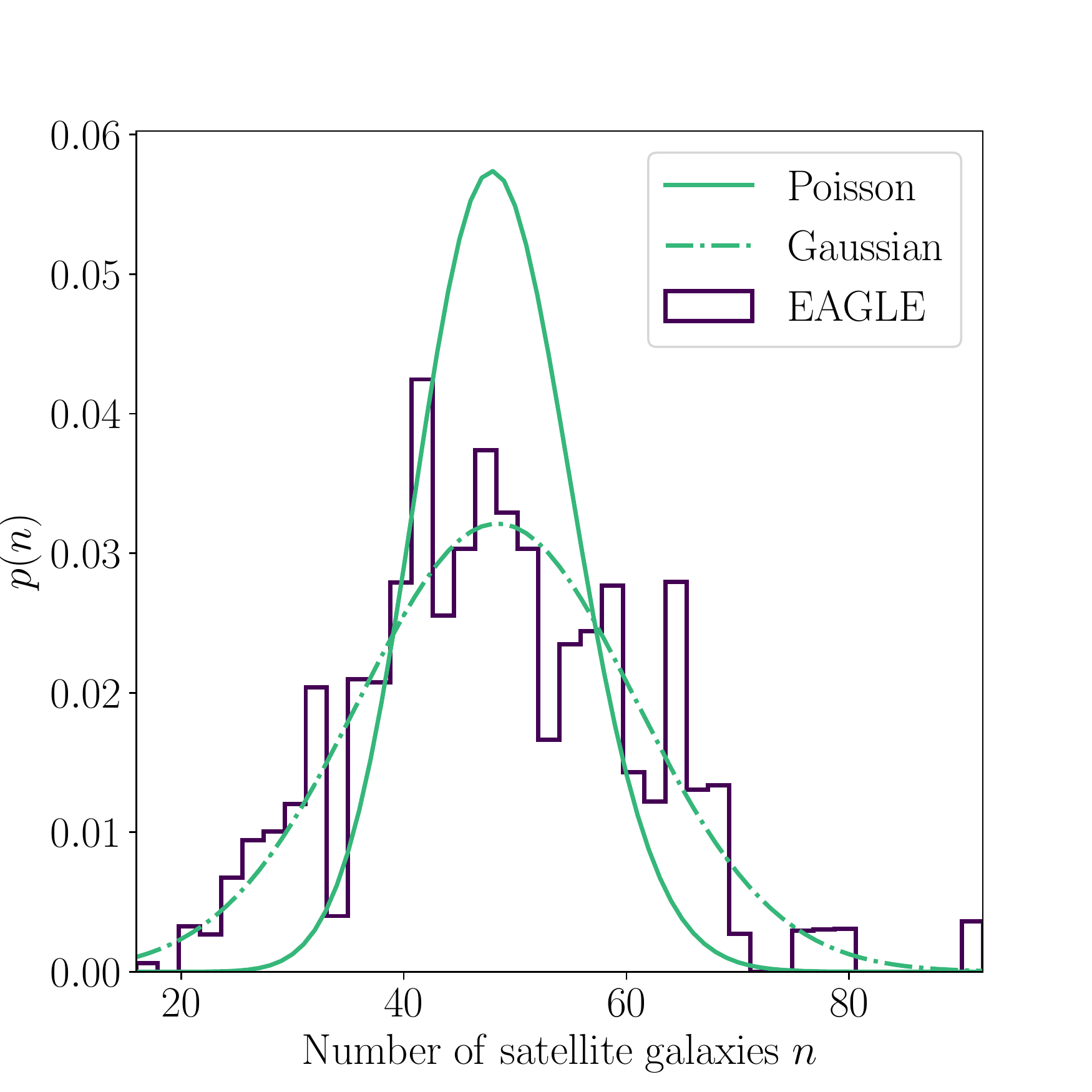}
 	\caption{Distribution of satellite galaxies in a halo of fixed mass within $12.0 < \log(M/M_{\odot}) < 12.2$ (histogram). This can be compared to a Poisson distribution with the same mean (solid curve) and a Gaussian distribution with the same mean and standard deviation as the data (dot-dashed curve).}
	\label{fig:eagle_poisson}
\end{figure}

\section{Discussion and conclusions}
\label{sec:discussion}

We have measured the projected galaxy clustering signal and galaxy-galaxy lensing signal for a sample of GAMA galaxies as a function of their stellar mass. In this analysis, we use the KiDS data covering 180 deg$^2$ of the sky \citep{Hildebrandt2016}, that fully overlaps with the three equatorial patches from the GAMA survey that we use to determine three stellar mass selected lens galaxy samples. We have combined our results to obtain the $\Gamma_{\text{gm}}(r_{\text{p}})$ bias function in order to unveil the hidden factors and origin of galaxy biasing in light of halo occupation models and the halo model, as presented in the theoretical work of \citet{Cacciato2012a}. We have used that formalism to fit to the data to constrain the parameters that contribute to the observed scale dependence of the galaxy bias, and see which parameters exactly carry information about the stochasticity and non-linearity of the galaxy bias, as observed. Due to the limited area covered by the both surveys, the covariance matrix used in this analysis was estimated using an analytical prescription, for which details can be found in Appendix \ref{sec:appendix}.

Our results show a clear trend that galaxy bias cannot be simply treated with a linear and/or deterministic approach. We find that the galaxy bias is inherently stochastic and non-linear due to the fact that satellite galaxies do not strictly follow a Poissonian distribution and that the spatial distribution of satellite galaxies also does not follow the NFW profile of the host dark matter halo. The main origin of the non-linearity of galaxy bias can be attributed to the fact that the central galaxy itself is heavily biased with respect to the dark matter halo in which it is residing. Those findings give additional support for the predictions presented by \citet{Cacciato2012a}, as their conclusions, based only on some fiducial model, are in line with our finding for a real subset of galaxies. We observe the same trends in the cosmological hydrodynamical simulation EAGLE, albeit out to smaller scales. We have also shown that the $\Gamma_{\text{gm}}(r_{\text{p}})$ bias function can, by itself, measure the properties of galaxy bias that would otherwise require the full knowledge of the $b_{\text{g}}(r_{\text{p}})$ and $\mathcal{R}_{\text{gm}}(r_{\text{p}})$ bias functions.

Our results are also in a broad agreement with recent findings of \citet{Gruen2017, Friedrich2017}, who used the density split statistics to measure the cosmological parameters in SDSS \citep{Rozo2016} and DES \citep{Drlica-Wagner2017} data, and as a byproduct, also the $b$ and $r$ functions directly (at angular scales around $20$ arcmin, which correspond to $3.5 - 7\, \text{Mpc}/h$ at redshifts of $0.2 - 4.5$). They find that the SDSS and DES data strongly prefer a stochastic bias with super-Poissonian behaviour. To obtain an independent measurement of galaxy bias and to further confirm our results we could use this method on our selection of galaxies, as well as the reconstruction method of \citet{Simon2017}. This work is, however, out of the scope of this paper.

Our findings show a remarkable wealth of information that halo occupation models are carrying in regard of understanding the nature of galaxy bias and its influence on cosmological analyses using the combination of galaxy-galaxy lensing and galaxy clustering. These results also show that the theoretical framework, as presented by \citet{Cacciato2012a}, is able to translate the constraints on galaxy biasing into constraints on galaxy formation and measurements of cosmological parameters.

As an extension of this work, we could fold in the cosmic shear measurements of the same sample of galaxies, and thus constrain the galaxy bias and the sources of non-linearity and stochasticity further. This would allow a direct measurement of all three bias functions [$\Gamma_{\text{gm}}(r_{\text{p}})$, $b_{\text{g}}(r_{\text{p}})$ and $\mathcal{R}_{\text{gm}}(r_{\text{p}})$], which could then be used directly in cosmological analyses. On the other hand, for a more detailed study of the HOD beyond those parameters that influence the galaxy bias, we could include the stellar mass (or luminosity) function in the joint fit. We leave such exercises open for future studies.

\section*{Acknowledgements}

We thank the anonymous referee for their very useful comments and suggestions. AD  would like to thank Marcello Cacciato for all the useful discussions, support and the hand written notes provided on the finer aspects of the theory used in this paper.

KK acknowledges support by the Alexander von Humboldt Foundation. HHo acknowledges support from Vici grant 639.043.512, financed by the Netherlands Organisation for Scientific Research (NWO). This work is supported by the Deutsche Forschungsgemeinschaft in the framework of the TR33 `The Dark Universe'. CH acknowledges support from the European Research Council under grant number 647112. HHi is supported by an Emmy Noether grant (No. Hi 1495/2-1) of the Deutsche Forschungsgemeinschaft. AA is supported by a LSSTC Data Science Fellowship. RN acknowledges support from the German Federal Ministry for Economic Affairs and Energy (BMWi) provided via DLR under project no. 50QE1103.

This research is based on data products from observations made with ESO Telescopes at the La Silla Paranal 
Observatory under programme IDs 177.A-3016, 177.A-3017 and 177.A-3018, and on data products produced 
by Target/OmegaCEN, INAF-OACN, INAF-OAPD and the KiDS production team, on behalf of the KiDS consortium.

GAMA is a joint European-Australasian project based around a spectroscopic campaign using the Anglo-Australian Telescope. 
The GAMA input catalogue is based on data taken from the Sloan Digital Sky Survey and the UKIRT Infrared Deep Sky Survey. 
Complementary imaging of the GAMA regions is being obtained by a number of independent survey programs including GALEX MIS, 
VST KiDS, VISTA VIKING, WISE, Herschel-ATLAS, GMRT and ASKAP providing UV to radio coverage. GAMA is funded by the 
STFC (UK), the ARC (Australia), the AAO, and the participating institutions. The GAMA website is \mbox{\url{http://www.gama-survey.org}}.

This work has made use of Python (\url{http://www.python.org}), including the packages \texttt{numpy} (\url{http://www.numpy.org}) 
and \texttt{scipy} (\url{http://www.scipy.org}). The halo model is built upon the \texttt{hmf} Python package by \citet{Murray2013}. 
Plots have been produced with \texttt{matplotlib} \citep{Hunter2007} and \texttt{corner.py} \citep{cornerplot}.

\emph{Author contributions:} All authors contributed to writing and development of this paper. The authorship list reflects the 
lead authors (AD, KK, HH, PS) followed by two alphabetical groups. The first alphabetical group includes those who are key 
contributors to both the scientific analysis and the data products. The second group covers those who have made a 
significant contribution either to the data products or to the scientific analysis.




\bibliographystyle{mnras}
\bibliography{library}



\appendix

\onecolumn

\section{Analytical covariance matrix}
\label{sec:appendix}

Here we present the expressions for the covariance of the auto-correlation and cross-correlation function of two observables, in our case specifically for the galaxy-galaxy, galaxy-mass auto-correlation functions and the cross-correlation function between the two. The expressions are an extension to the Gaussian part of the covariance as presented in \citet{Singh2016} and include the non-Gaussian terms and the super-sample covariance terms that are by themselves an extension \citep{Krause2016} to the non-Gaussian terms as previously described for cosmic shear only by \citet{Takada2013, Li2014}. We follow \citet{Singh2016, Marian2015} for the Gaussian terms, but excluding the additional contributions that arise due to not subtracting the signal obtained using random positions on the sky, as in our analysis this is performed during the signal extraction. 

In general, the covariance matrix between two observables can be written as:
\begin{equation}
\text{Cov}(X,Y) = \text{Cov}^{\text{G}}(X, Y) + \text{Cov}^{\text{NG}}(X, Y) + \text{Cov}^{\text{SSC}}(X, Y) \,,
\end{equation}
where $X$ and $Y$ are either $w_{\text{p}}(r_{\text{p}})$ or $\Delta \Sigma_{\text{gm}}(r_{\text{p}})$, and the G stands for the Gaussian term, NG for the non-Gaussian term and SSC stands for the contributions from the super-sample covariance. Furthermore, following \citet[][starting with equation A18]{Singh2016} and \citet[][using the derivations in their Appendix]{Marian2015}, the Gaussian terms for each auto-correlation or cross-correlation can be written as (where indices $i, j$ stand for individual projected radial bins and indices $n,m$ stand for individual galaxy sample bins):
\begin{align}\label{cov_esd}
\text{Cov}^{\text{G}}[w_{\text{p}}^{n} \,, w_{\text{p}}^{m}](r_{\text{p},i} \,, r_{\text{p},j}) &= 2 \frac{\mathcal{A}_{W, 2} ( r_{\text{p},i}, r_{\text{p},j} ) }{ \mathcal{A}_{W, 1} (r_{\text{p},i})\, \mathcal{A}_{W, 1} (r_{\text{p},j}) } \int {\text{d}k\,k \over 2 \pi} J_{0}(kr_{\text{p},i})\,  J_{0}(kr_{\text{p},j}) \left( P_{\text{gg}}^{n}(k)  + \delta_{nm}{1 \over \overline{n}_{\text{g}}^{n}} \right) \left(P_{\text{gg}}^{m}(k)  + \delta_{nm}{1 \over \overline{n}_{\text{g}}^{m}}\right)   \,, 
\end{align}
\begin{align}\label{cov_wp}
&\text{Cov}^{\text{G}}[\Delta \Sigma_{\text{gm}}^{n} \,, \Delta \Sigma_{\text{gm}}^{m}](r_{\text{p},i} \,, r_{\text{p},j}) = \overline{\rho}_{\text{m}}^{2} \frac{\mathcal{A}_{W, 2} ( r_{\text{p},i}, r_{\text{p},j} ) }{ \mathcal{A}_{W, 1} (r_{\text{p},i})\, \mathcal{A}_{W, 1} (r_{\text{p},j}) } \int {\text{d}k\,k \over 2 \pi} J_{2}(kr_{\text{p},i})\,  J_{2}(kr_{\text{p},j})  \left(P_{\text{gg}}^{n}(k)  + \delta_{nm}{1 \over \overline{n}_{\text{g}}^{n}}\right) \left(P_{\text{mm}}^{m}(k) + \delta_{nm} \frac{1}{\overline{n}_{\gamma}} \right) \nonumber \\
&\qquad \qquad \qquad \qquad \qquad \qquad \  + \overline{\rho}_{\text{m}}^{2} \frac{ \mathcal{A}_{W, 2} ( r_{\text{p},i}, r_{\text{p},j} ) }{ \mathcal{A}_{W, 1} (r_{\text{p},i})\, \mathcal{A}_{W, 1} (r_{\text{p},j}) } \int {\text{d}k\,k \over 2 \pi} J_{2}(kr_{\text{p},i})\,  J_{2}(kr_{\text{p},j}) \,  P_{\text{gm}}^{n}(k) \, P_{\text{gm}}^{m}(k)  \,, 
\end{align}
\begin{align}\label{cov_wp_esd}
\text{Cov}^{\text{G}}[w_{\text{p}}^{n} \,, \Delta \Sigma_{\text{p}}^{m}](r_{\text{p},i} \,, r_{\text{p},j}) &=  \overline{\rho}_{\text{m}} \frac{ \mathcal{A}_{W, 2} ( r_{\text{p},i}, r_{\text{p},j} ) }{ \mathcal{A}_{W, 1} (r_{\text{p},i})\, \mathcal{A}_{W, 1} (r_{\text{p},j}) } \int {\text{d}k\,k \over 2 \pi} J_{0}(kr_{\text{p},i})\,  J_{2}(kr_{\text{p},j}) \left[ \left(P_{\text{gg}}^{n}(k)  + \delta_{nm}{1 \over \overline{n}_{\text{g}}^{n}} \right)  \, P_{\text{gm}}^{n}  + \left(P_{\text{gg}}^{m}(k)  + \delta_{nm}{1 \over \overline{n}_{\text{g}}^{m}} \right)  \, P_{\text{gm}}^{m} \right] \,,
\end{align}
where
\begin{equation}
\mathcal{A}_{W,1} (r_{\text{p}}) = \int {\text{d}k\,k \over 2 \pi} J_{0}(kr_{\text{p}})\, \widetilde{W}^{2}(k)
\end{equation}
and
\begin{equation}
\mathcal{A}_{W,2} ( r_{\text{p},i}, r_{\text{p},j} )  = \int {\text{d}k\,k \over 2 \pi} J_{0}(kr_{\text{p},i})\, J_{0}(kr_{\text{p},j})\, \widetilde{W}^{2}(k)
\end{equation}
are the pre-factors arising from the survey geometry, with $J_{n}$ being Bessel function of the $n$-th kind, $\widetilde{W}(k)$ is the window function defined in equation (\ref{eq:window}), $\delta_{nm}$ is the Kronecker delta symbol, $\overline{n}_{\text{g}}$ the number density of galaxies in the bin $n$, $P_{\text{xy}}(k)$ are individual power spectra as defined in Section \ref{sec:halomodel_ingredients}, and the $\overline{n}_{\gamma}$ is the shape noise given by \citep[see also][equation C2]{Marian2015}:
\begin{equation}
\frac{1}{\overline{n}_{\gamma}} = \frac{\sigma_{\gamma}^{2}}{\overline{n}_{\text{s}}} \, \frac{\Sigma_{\text{cr,com}}^{2}}{\overline{\rho}_{\text{m}}^{2}} \, \frac{D^{2}(z_{\text{l}})}{2\Pi_{\text{max}}} \,,
\end{equation}
where $\sigma_{\gamma}$ is the shape variance of the sources used in the analysis, the $\overline{n}_{\text{s}}$ is the source density given by \citet{Hildebrandt2016} in gal/arcmin$^{2}$ (converted to radians), the $D(z_{\text{l}})$ is the angular diameter distance at $z_{\text{l}}$ and $\Pi_{\text{max}}$ is the projection length used throughout this work. The value $\overline{n}_{\text{g}}$ and $\sigma_{\gamma}$ are measured from the lens and source galaxies used in our $3$ samples, respectively. We assume a circular survey geometry with a window function:
\begin{equation}\label{eq:window}
\widetilde{W}(k) = 2\pi R^{2} \frac{J_{1}(kR)}{kR}\,,
\end{equation}
where $R$ is the radius of the circular window with area covering $180$ deg$^2$. The $\Sigma_{\text{cr,com}}$ is calculated using the same prescription as defined in equation (\ref{eq:crit_effective}). To project our covariance matrices, we use the Limber approximation as demonstrated by \citet{Marian2015}, using the slightly more accurate survey area normalisation by \citet{Singh2016}.

The super-sample covariance terms are given by the following expressions: 
\begin{equation}\label{eq:ssc_1}
\text{Cov}^{\text{SSC}}[w_{\text{p}}^{n} \,, w_{\text{p}}^{m}](r_{\text{p},i} \,, r_{\text{p},j}) = \frac{\mathcal{A}_{W, 2} ( r_{\text{p},i}, r_{\text{p},j} ) }{ \mathcal{A}_{W, 1} (r_{\text{p},i})\, \mathcal{A}_{W, 1} (r_{\text{p},j}) } \int {\text{d}k\,k \over 2 \pi} J_{0}(kr_{\text{p},i})\,  J_{0}(kr_{\text{p},j}) \frac{\partial P_{\text{gg}}^{n}(k)}{\partial \delta_{\text{b}}}\, \frac{\partial P_{\text{gg}}^{m}(k)}{\partial \delta_{\text{b}}} \, \sigma_{\text{s}, nm}^{2} \,,
\end{equation}
\begin{equation}\label{eq:ssc_2}
\text{Cov}^{\text{SSC}}[\Delta \Sigma_{\text{gm}}^{n} \,, \Delta \Sigma_{\text{gm}}^{m}](r_{\text{p},i} \,, r_{\text{p},j}) = \overline{\rho}_{\text{m}}^{2} \frac{\mathcal{A}_{W, 2} ( r_{\text{p},i}, r_{\text{p},j} ) }{ \mathcal{A}_{W, 1} (r_{\text{p},i})\, \mathcal{A}_{W, 1} (r_{\text{p},j}) } \int {\text{d}k\,k \over 2 \pi} J_{2}(kr_{\text{p},i})\,  J_{2}(kr_{\text{p},j}) \frac{\partial P_{\text{gm}}^{n}(k)}{\partial \delta_{\text{b}}}\, \frac{\partial P_{\text{gm}}^{m}(k)}{\partial \delta_{\text{b}}} \, \sigma_{\text{s}, nm}^{2} \,,
\end{equation}
\begin{equation}\label{eq:ssc_3}
\text{Cov}^{\text{SSC}}[w_{\text{p}}^{n} \,, \Delta \Sigma_{\text{p}}^{m}](r_{\text{p},i} \,, r_{\text{p},j}) = \overline{\rho}_{\text{m}} \frac{\mathcal{A}_{W, 2} ( r_{\text{p},i}, r_{\text{p},j} ) }{ \mathcal{A}_{W, 1} (r_{\text{p},i})\, \mathcal{A}_{W, 1} (r_{\text{p},j}) } \int {\text{d}k\,k \over 2 \pi} J_{0}(kr_{\text{p},i})\,  J_{2}(kr_{\text{p},j}) \frac{\partial P_{\text{gg}}^{n}(k)}{\partial \delta_{\text{b}}}\, \frac{\partial P_{\text{gm}}^{m}(k)}{\partial \delta_{\text{b}}} \, \sigma_{\text{s}, nm}^{2} \,,
\end{equation}
where the responses $\frac{\partial P_{\text{xy}}(k)}{\partial \delta_{\text{b}}}$ are given by the following equations [following galaxy-matter and galaxy-galaxy extensions to the matter-matter only responses in \citet{Takada2013, Li2014} as derived by \citet{Krause2016}]:
\begin{equation}
\frac{\partial P_{\text{gg}}(k)}{\partial \delta_{\text{b}}} = \left( \frac{68}{21} - \frac{1}{3}\frac{\text{d}\ln k^{3} P_{\text{lin}}(k)}{\text{d} \ln k} \right) I_{\text{g}}^{1}(k) I_{\text{g}}^{1}(k) P_{\text{lin}}(k) + I_{\text{gg}}^{1}(k,k) - 2 b_{\text{g},1} P_{\text{gg}}(k) \,,
\end{equation}
\begin{equation}
\frac{\partial P_{\text{gm}}(k)}{\partial \delta_{\text{b}}} = \left( \frac{68}{21} - \frac{1}{3}\frac{\text{d}\ln k^{3} P_{\text{lin}}(k)}{\text{d} \ln k} \right) I_{\text{g}}^{1}(k) I_{\text{m}}^{1}(k) P_{\text{lin}}(k) + I_{\text{gm}}^{1}(k,k) -  b_{\text{g},1} P_{\text{gm}}(k) \,,
\end{equation} 
\begin{equation}
\frac{\partial P_{\text{mm}}(k)}{\partial \delta_{\text{b}}} = \left( \frac{68}{21} - \frac{1}{3}\frac{\text{d}\ln k^{3} P_{\text{lin}}(k)}{\text{d} \ln k} \right) I_{\text{m}}^{1}(k) I_{\text{m}}^{1}(k) P_{\text{lin}}(k) + I_{\text{mm}}^{1}(k,k)  \,,
\end{equation}
where we have introduced the halo model integrals $I_{\text{x}}^{\alpha}(k)$ and $I_{\text{xy}}^{\alpha}(k, k')$ as:
\begin{equation}\label{eq:I_1}
I_{\text{x}}^{\alpha}(k) = \int \mathcal{H}_{\text{x}}(k, M) \, b_{i, \alpha}\, n(M) \, \text{d} M\,,
\end{equation}
\begin{equation}\label{eq:I_2}
I_{\text{xy}}^{\alpha}(k, k') = \int \mathcal{H}_{\text{x}}(k,M)\, \mathcal{H}_{\text{y}}(k',M) \, b_{i, \alpha}\, n(M) \, \text{d} M\,.
\end{equation}
In the case when `x' equals `g', $\mathcal{H}_{\text{g}}(k,M)$ is the sum of the $\mathcal{H}_{\text{c}}(k,M)$ and $\mathcal{H}_{\text{s}}(k,M)$ functions. Bias functions $b_{i, \alpha}$ are either $0$ if $\alpha = 0$,  1 if $i = \text{m}$ and $\alpha = 1$, and $ \int \text{d} M \, \mathcal{H}_{\text{g}}(k, M) \, b_{\text{h}}(M)\, n(M) / \overline{n}_{\text{g}}   $  if $i = \text{g}$ and $\alpha = 1$. Note that these subscripts are not related to the ones used in Section \ref{sec:halomodel_ingredients}. In the equations (\ref{eq:ssc_1}), (\ref{eq:ssc_2}) and (\ref{eq:ssc_3}), we have also used the survey variance defined as:
\begin{equation}
\sigma_{\text{s}}^{2} = \frac{1}{\mathcal{A}_{W, 1} (R)} \int {\text{d}k\,k^{2} \over 2 \pi^{2}} \widetilde{W}^{2}(k) P_{\text{lin}}(k)\,.
\end{equation}

The connected, non-Gaussian terms of the covariance matrix can be written as \citep[again following the extension to the matter-matter only derivation by][]{Krause2016} \citep[see also][]{Cooray2002, Takada2008}:
\begin{equation}
\text{Cov}^{\text{NG}}[w_{\text{p}}^{n} \,, w_{\text{p}}^{m}](r_{\text{p},i} \,, r_{\text{p},j}) = \frac{\mathcal{A}_{W, 2} ( r_{\text{p},i}, r_{\text{p},j} ) }{ \mathcal{A}_{W, 1} (r_{\text{p},i})\, \mathcal{A}_{W, 1} (r_{\text{p},j}) } \int {\text{d}k_{i}\,k_{i} \over \sqrt{2 \pi}} \int {\text{d}k_{j}\,k_{j} \over \sqrt{2 \pi}} J_{0}(k_{i}r_{\text{p},i})\,  J_{0}(k_{j}r_{\text{p},j}) \cdot T_{\text{gggg}}^{nm}(k_{i}, -k_{i}, k_{j}, -k_{j}) \,,
\end{equation}
\begin{equation}
\text{Cov}^{\text{NG}}[\Delta \Sigma_{\text{gm}}^{n} \,, \Delta \Sigma_{\text{gm}}^{m}](r_{\text{p},i} \,, r_{\text{p},j}) = \overline{\rho}_{\text{m}}^{2} \frac{\mathcal{A}_{W, 2} ( r_{\text{p},i}, r_{\text{p},j} ) }{ \mathcal{A}_{W, 1} (r_{\text{p},i})\, \mathcal{A}_{W, 1} (r_{\text{p},j}) } \int {\text{d}k_{i}\,k_{i} \over \sqrt{2 \pi}} \int {\text{d}k_{j}\,k_{j} \over \sqrt{2 \pi}} J_{2}(k_{i}r_{\text{p},i})\,  J_{2}(k_{j}r_{\text{p},j})\cdot T_{\text{gmgm}}^{nm}(k_{i}, -k_{i}, k_{j}, -k_{j}) \,,
\end{equation}
\begin{equation}
\text{Cov}^{\text{NG}}[w_{\text{p}}^{n} \,, \Delta \Sigma_{\text{p}}^{m}](r_{\text{p},i} \,, r_{\text{p},j}) = \overline{\rho}_{\text{m}} \frac{\mathcal{A}_{W, 2} ( r_{\text{p},i}, r_{\text{p},j} ) }{ \mathcal{A}_{W, 1} (r_{\text{p},i})\, \mathcal{A}_{W, 1} (r_{\text{p},j}) } \int {\text{d}k_{i}\,k_{i} \over \sqrt{2 \pi}} \int {\text{d}k_{j}\,k_{j} \over \sqrt{2 \pi}} J_{0}(k_{i}r_{\text{p},i})\,  J_{2}(k_{j}r_{\text{p},j})\cdot  T_{\text{gggm}}^{nm}(k_{i}, -k_{i}, k_{j}, -k_{j}) \,,
\end{equation}
where the individual $T_{\text{xyzw}}$ terms are given by the combination of 1-halo matter trispectrum and $(2+3+4)$-halo matter trispectrum as:
\begin{equation}
T_{\text{xyzw}}^{nm}(k_{i}, -k_{i}, k_{j}, -k_{j}) = (b_{\text{x}}b_{\text{y}}b_{\text{z}}b_{\text{w}})^{nm}\, T_{\text{m}}^{2\text{h} + 3\text{h} + 4\text{h}} (k_{i}, -k_{i}, k_{j}, -k_{j}) + T_{\text{xyzw}}^{1\text{h}, nm}(k_{i}, k_{i}, k_{j}, k_{j}) \,,
\end{equation}
where $b_{\text{x}}$ is the same bias used in equations (\ref{eq:I_1}) and (\ref{eq:I_2}). The 1-halo matter trispectrum is given by the following integral over halo model building blocks:
\begin{equation}
T_{\text{xyzw}}^{1\text{h}, nm}(k_{i}, k_{i}, k_{j}, k_{j}) = \int \text{d} M \, n(M) \mathcal{H}_{\text{x}}^{n}(k_{i},M) \mathcal{H}_{\text{y}}^{n}(k_{i},M) \mathcal{H}_{\text{z}}^{m}(k_{j},M) \mathcal{H}_{\text{w}}^{m}(k_{j},M)\,,
\end{equation}
and the $(2+3+4)$-halo matter trispectrum $T_{\text{m}}^{2\text{h} + 3\text{h} + 4\text{h}}$ is calculated according to the estimate presented in \citet{Pielorz2010}. Additionally, the resulting covariance matrix is bin averaged according to the same radial binning scheme as used with our data. The resulting covariance matrix obtained using the analytical prescription presented here can be seen in Figure \ref{fig:cov_analytical_full} (shown as a correlation matrix), which shows the auto-correlation between the 3 bins for the lensing signal, clustering signal and the cross-correlation between the two observables in the off-diagonal $3\times 3$ blocks. Individual combinations between all the bins are marked above the corresponding block matrices.

\begin{figure*}\label{fig:cov_analytical_full}
	\includegraphics[width=\textwidth]{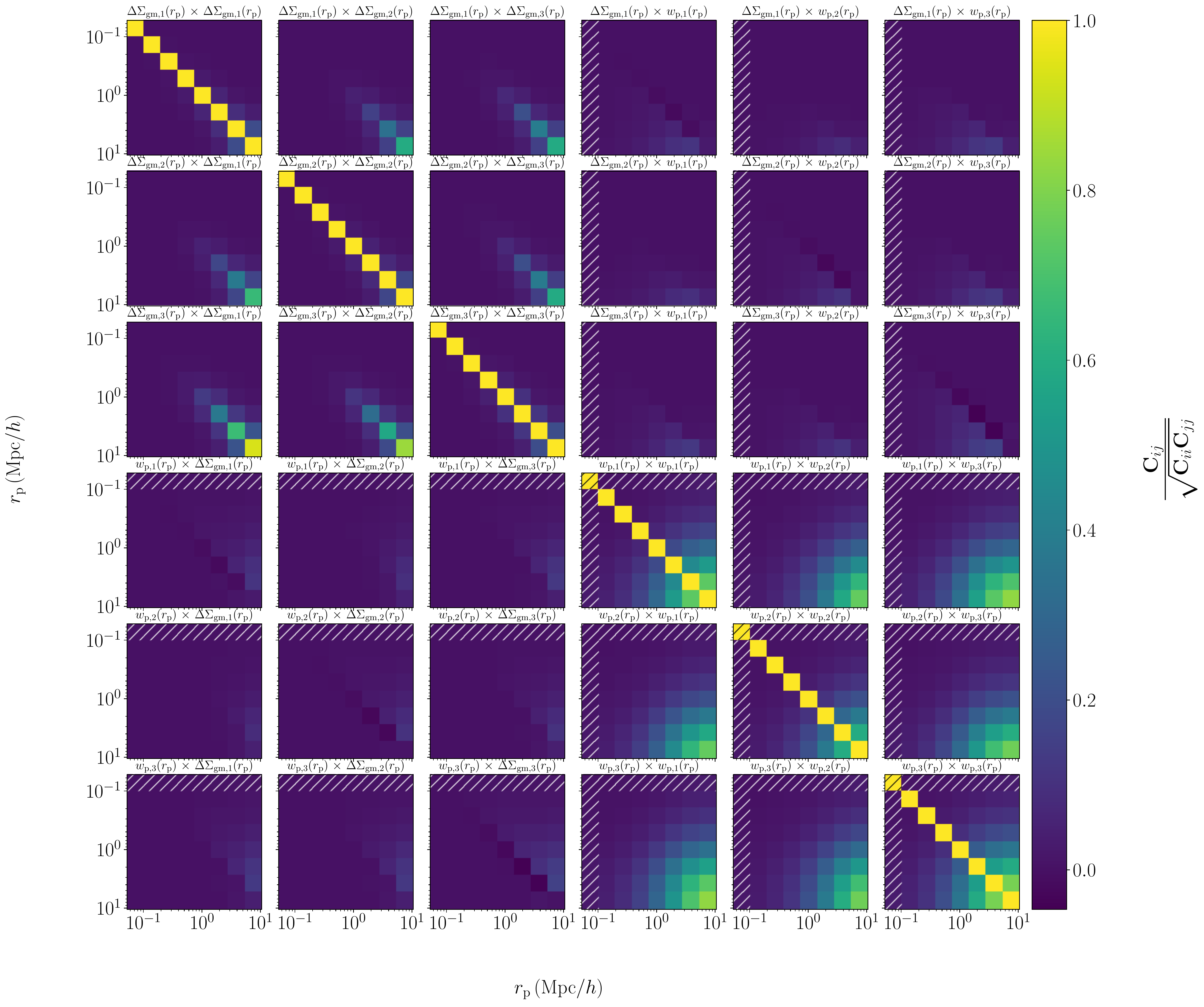}
 	\caption{The full analytical correlation matrix for the lensing and clustering signals and their cross terms. Individual combinations between all the bins are marked above the corresponding block matrices, with indices 1,2 and 3 corresponding to the stellar mass bins as defined in Table \ref{tab:sample_properties}. We do not use the covariance estimates in the hatched areas in our fit, as the clustering measurements are affected by the blending on these scales.}
\end{figure*}

\section{Full posterior distributions}
\label{sec:appendix2}

In Figure \ref{fig:triangle} we show the full posterior probability distribution for all fitted parameters in our MCMC fit as discussed in Section \ref{sec:data}.

\begin{figure*}
	\includegraphics[width=\textwidth]{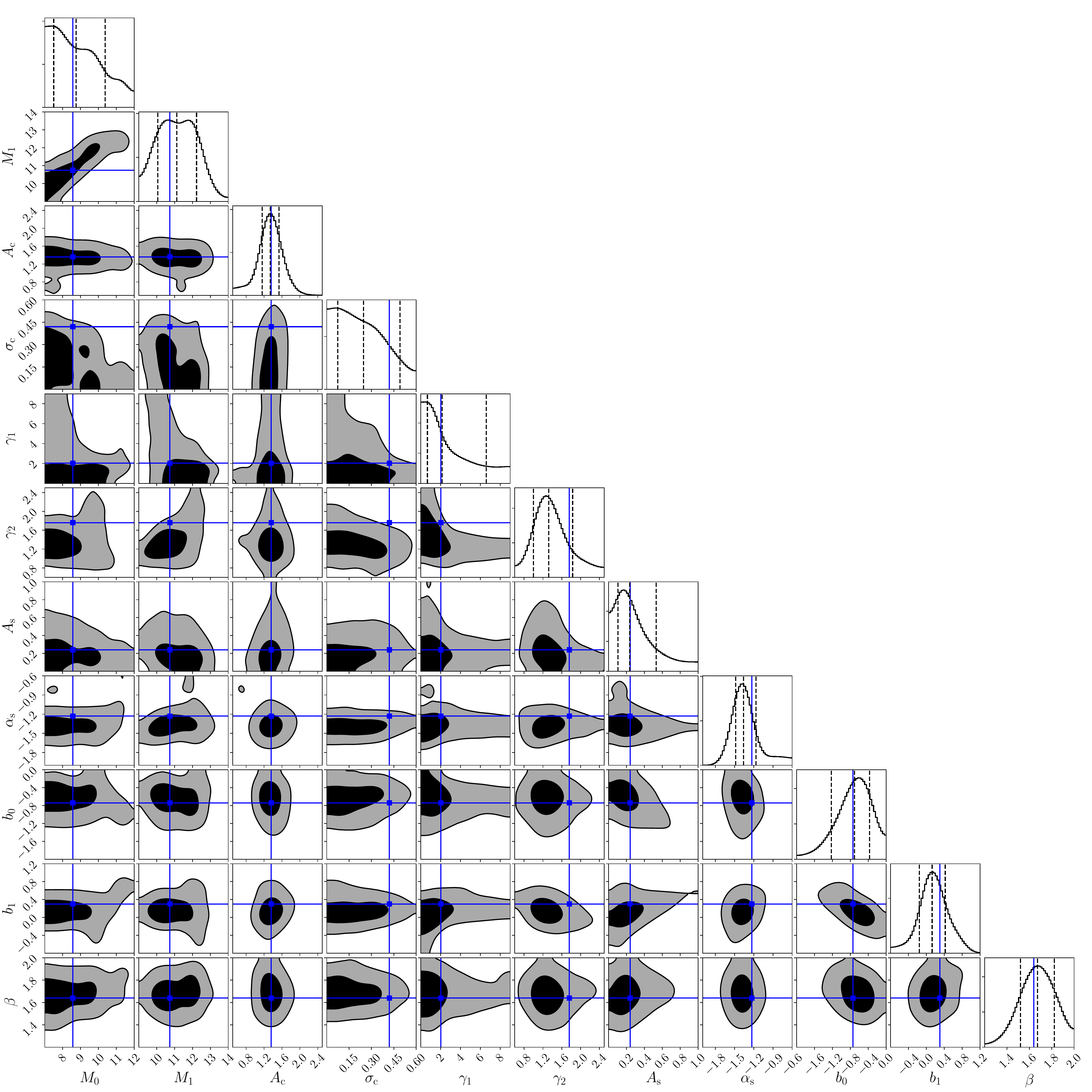}
 	\caption{The full posterior distributions of the halo model parameters (where the priors are listed in Table \ref{tab:results}). The contours indicate $1\sigma$ and $2\sigma$ confidence regions.}
	\label{fig:triangle}
\end{figure*}

\onecolumn

\section{Relation between the galaxy-galaxy lensing signal and the galaxy-matter cross-correlation function}
\label{sec:appendix2}

In this appendix, we provide a step-by-step derivation of the relation
between the galaxy-galaxy lensing signal and the galaxy-matter
cross-correlation function. As a side-product, we motivate the two
different definitions of the critical surface mass density that are
used in this field. Finally, we compare our results with those in some
recent papers, pointing out differences, and discussing their
implications. Since the results of this appendix apply to several papers, 
we choose to use a slightly more explicit notation
here in comparison to the rest of this paper.

\subsection{Derivation}
\label{sec:appendix2deriv}
The equivalent weak lensing convergence $\kappa$ for a
three-dimensional mass distribution characterized by the fractional
density contrast $\delta$, for sources at comoving distance $\chi_{\rm
  s}$, is given by \citep[e.g.,][]{Bartelmann2001, Schneider2006}
\begin{equation}
\kappa(\boldsymbol{\theta})={3 H_0^2\Omega_{\rm m}\over 2 c^2}
\int_0^{\chi_{\rm s}}\mathrm{d}\chi\;
{\chi\rund{\chi_{\rm s}-\chi}\over \chi_{\rm s}\,a(\chi)}
\,\delta(\chi\boldsymbol{\theta},\chi) 
=\overline{\rho}_{\rm m}\,{4\pi G\over c^2}  
\int_0^{\chi_{\rm s}}\mathrm{d}\chi\;
{\chi\rund{\chi_{\rm s}-\chi}\over \chi_{\rm s}\,a(\chi)}
\,\delta(\chi\boldsymbol{\theta},\chi) \;,
\label{eq:1}
\end{equation}
where we assumed for notational simplicity a spatially flat
cosmological model.
Here, $\overline{\rho}_{\rm m}$ is the current mean matter density in the
Universe, and we used the relation between mass density and density parameter in
the second step, i.e., $\overline{\rho}_{\rm m}=3 H_0^2 \Omega_{\rm m}/(8\pi
G)$.  The relation (\ref{eq:1}) is valid in the framework
of the Born approximation and by neglecting lens-lens coupling \citep[see, e.g.][for the impact of
these effects]{Hilbert2009, Krause2010}. 

Let $\delta_{\rm g}$ be the three-dimensional fractional density
contrast of galaxies of a given type. Their fractional density contrast
on the sky, $\kappa_{\rm g}(\boldsymbol{\theta})=[n(\boldsymbol{\theta})-\overline{n}]/\overline{n}$,
with $\overline{n}$ being the mean number density,
is related to $\delta_{\rm g}$ by
\begin{equation}
\kappa_{\rm g}(\boldsymbol{\theta})=\int\mathrm{d}\chi\;p_{\rm f}(\chi)\,\delta_{\rm g}
(\chi\boldsymbol{\theta},\chi)\;,
\label{eq:2}
\end{equation}
where $p_{\rm f}(\chi)$ is the probability distribution of the
selected `foreground' galaxy population in comoving distance,
equivalent to a redshift probability distribution. For the
following we will assume that this distribution is a very narrow one
around redshift $z_{\rm l}$, and thus approximate $p_{\rm f}(\chi)=\delta_{\rm
  D}(\chi-\chi_{\rm l})$. We assume throughout that $\chi_{\rm
  l}<\chi_{\rm s}$. The correlator between $\kappa$ and
$\kappa_{\rm g}$ then becomes
\begin{align}
\ave{\kappa_{\rm g}(\boldsymbol{\theta})\,\kappa(\boldsymbol{\theta}+\boldsymbol{\vartheta})}
&=\overline{\rho}_{\rm m}\,{4\pi G\over c^2} 
\int_0^{\chi_{\rm s}}\mathrm{d}\chi\;
{\chi\rund{\chi_{\rm s}-\chi}\over \chi_{\rm s}\,a(\chi)}
\int \mathrm{d}\chi'\;\delta_{\rm D}(\chi'-\chi_{\rm l})
\ave{\delta_{\rm g}\rund{\chi'\boldsymbol{\theta},\chi'}\;
\delta\eck{\chi\rund{\boldsymbol{\theta}+\boldsymbol{\vartheta}},\chi}} \nonumber \\
&= \overline{\rho}_{\rm m}\,{4\pi G\over c^2} 
\int_0^{\chi_{\rm s}}\mathrm{d}\chi\;
{\chi\rund{\chi_{\rm s}-\chi}\over \chi_{\rm s}\,a(\chi)}
\ave{\delta_{\rm g}\rund{\chi_{\rm l}\boldsymbol{\theta},\chi_{\rm l}}\;
\delta\eck{\chi\rund{\boldsymbol{\theta}+\boldsymbol{\vartheta}},\chi}} \;.
\label{eq:3}
\end{align}
Since the correlator is significantly non-zero only over a small
interval in $\chi$ around $\chi_{\rm l}$, the prefactor in the
integrand can be considered to be constant over this interval and taken
out of the integral. This yields
\begin{align}
\ave{\kappa_{\rm g}(\boldsymbol{\theta})\,\kappa(\boldsymbol{\theta}+\boldsymbol{\vartheta})}
&={4\pi G\over c^2}\,{\chi_{\rm l}
 \rund{\chi_{\rm s}-\chi_{\rm l}}\over
  \chi_{\rm s}\,a(\chi_{\rm l})}\;
\overline{\rho}_{\rm m}
\int_0^{\chi_{\rm s}}\mathrm{d}\chi\; \xi_{\rm gm}
\rund{\sqrt{\chi_{\rm l}^2|\boldsymbol{\vartheta}|^2
+(\chi-\chi_{\rm l})^2}}  \nonumber \\
&= \Sigma_{\rm cr,com}^{-1}\; \Sigma_{\rm com}(\chi_{\rm
  l}|\boldsymbol{\theta}|) \;,
\label{eq:4}
\end{align}
where the
galaxy-matter cross-correlation function $\xi_{\rm gm}$ (at fixed redshift
$z_{\rm l}$) is defined through
\begin{equation}
\ave{\delta(\vc x)\,\delta_{\rm g}(\vc x+\vc y)}=\xi_{\rm gm}(|\vc
  y|)\;,
\label{eq:5}
\end{equation}
in which $\vc x$ and $\vc y$ are comoving spatial vectors, and 
the sole dependence on $|\vc y|$ is due to the assumed
homogeneity and isotropy of the density fields in the Universe.
Furthermore,
we have defined the comoving critical surface
mass density $\Sigma_{\rm cr,com}$ through
\begin{equation}
\Sigma_{\rm cr,com}^{-1}={4\pi G\over c^2}\,{\chi_{\rm l}
 \rund{\chi_{\rm s}-\chi_{\rm l}}\over
  \chi_{\rm s}\,a(\chi_{\rm l})} \,{\rm H}(\chi_{\rm s}-\chi_{\rm l}) \;,
\label{eq:6}
\end{equation}
with ${\rm H}(x)$ being the Heaviside unit step function,\footnote{The
  corresponding expression for a general curvature parameter reads
\[
\Sigma_{\rm cr,com}^{-1}={4\pi G\over c^2}\,{f_K(\chi_{\rm l})
 f_K\rund{\chi_{\rm s}-\chi_{\rm l}}\over
  f_K(\chi_{\rm s})\,a(\chi_{\rm l})} \,{\rm H}(\chi_{\rm s}-\chi_{\rm l}) \;,
\]
where $f_K(\chi)$ is the comoving angular-diameter distance to a
comoving distance $\chi$, and either the identity for spatially flat
models, or a sin or sinh function for closed or open models, respectively.}
and the comoving surface mass density as
\begin{equation}
\Sigma_{\rm com}(R_{\rm com})=
\overline{\rho}_{\rm m}
\int_0^{\chi_{\rm s}}\mathrm{d}\chi\; \xi_{\rm gm}
\rund{\sqrt{R_{\rm com}^2
+(\chi-\chi_{\rm l})^2}} \;,
\label{eq:7}
\end{equation}
as a function of the comoving projected separation $R_{\rm com}$.
In this paper, $\Sigma_{\rm cr,com}$ is termed $\Sigma_{\rm crit}$ --
see equation (\ref{eq:SigmaCrit}), and $R_{\rm com}$ and $\Sigma_{\rm com}$ are called
$r_{\rm p}$ and $\Sigma_{\rm gm}$ -- see equation (\ref{eq:xyprojdef}).

The interpretation
of equation (\ref{eq:4}) is then that $\overline{\rho}_{\rm m} \xi_{\rm gm}$ is the
average comoving overdensity of matter around galaxies, caused by the
correlation between them, and that the integral over comoving distance
then yields the comoving surface mass density of this excess
matter. The corresponding convergence is then obtained by scaling with
the comoving critical surface mass density $\Sigma_{\rm cr,com}$.

There is another form in which equation (\ref{eq:4}) can be written by
rearranging factors of $a(\chi_{\rm l})$, namely
\begin{align}
\ave{\kappa_{\rm g}(\boldsymbol{\theta})\,\kappa(\boldsymbol{\theta}+\boldsymbol{\vartheta})}
&={4\pi G\over c^2}\,{\chi_{\rm l}
 \rund{\chi_{\rm s}-\chi_{\rm l}}\,a(\chi_{\rm l})\over
  \chi_{\rm s}}\;
\overline{\rho}_{\rm m}\,a^{-2}(\chi_{\rm l})
\int_0^{\chi_{\rm s}}\mathrm{d}\chi\; \xi_{\rm gm}
\rund{\sqrt{\chi_{\rm l}^2|\boldsymbol{\vartheta}|^2
+(\chi-\chi_{\rm l})^2}} \nonumber \\
&=: \Sigma_{\rm cr}^{-1}\; \Sigma(\chi_{\rm l}|\boldsymbol{\theta}|) \;,
\label{eq:8}
\end{align}
where we defined the (proper) critical surface mass
density
$\Sigma_{\rm cr}$ through
\begin{equation}
\Sigma_{\rm cr}^{-1}={4\pi G\over c^2}\,{\chi_{\rm l}
 \rund{\chi_{\rm s}-\chi_{\rm l}}\,a(\chi_{\rm l})\over
  \chi_{\rm s}}\,{\rm H}(\chi_{\rm s}-\chi_{\rm d})
={4\pi G\over c^2}\,{D_{\rm l} D_{\rm ls}\over D_{\rm
    s}}\;,
\label{eq:9}
\end{equation}
and in the last step we introduced the angular-diameter distances
$D_{\rm l}=D(0,z_{\rm l})$, $D_{\rm s}=D(0,z_{\rm s})$ and $D_{\rm
  ds}=D(z_{\rm l},z_{\rm s})$, with
$D(z_1,z_2)=a(z_2)\eck{\chi(z_2)-\chi(z_1)}\;{\rm H}(z_2-z_1)$ being 
the angular-diameter distance of a source at redshift $z_2$ as seen
from an observer at redshift $z_1$.\footnote{For a model with free
  curvature,
  $D(z_1,z_2)=a(z_2)\wave{f_K\eck{\chi(z_2)-\chi(z_1)}}\;{\rm
    H}(z_2-z_1)$.} Furthermore, 
\begin{equation}
\Sigma(R_{\rm com})=\rho_{\rm m}(\chi_{\rm l})
\int_0^{\chi_{\rm s}}\mathrm{d}\chi\,a(\chi)\; \xi_{\rm gm}
\rund{\sqrt{R_{\rm com}^2
+(\chi-\chi_{\rm l})^2}} \;.
\label{eq:10}
\end{equation}
We note that, due to the assumed localized nature of the
correlation function, we could write $a(\chi)$ into the integrand in
equation (\ref{eq:10}). 
The interpretation of equation (\ref{eq:8})  is now that the
(proper) overdensity around galaxies caused by the galaxy-matter
cross-correlation, $\rho_{\rm m}(\chi_{\rm l}) \xi_{\rm gm}=\overline{\rho}_{\rm
  m}(1+z_{\rm l})^3 \xi_{\rm gm}$, is integrated along the l.o.s. in
proper coordinates, $\mathrm{d} r_{\rm prop}=a\,\mathrm{d}\chi$, and the resulting
(proper) surface mass density is scaled by the critical surface mass
density $\Sigma_{\rm cr}$. We note that the argument of the (proper)
surface mass density $\Sigma$ is a comoving transverse separation,
since the correlation function is a function of comoving
separation. 

The relation between the two different
equations (\ref{eq:6}) and (\ref{eq:9}) of the critical surface mass
density is 
\begin{equation}
\Sigma_{\rm cr,com}=a^2(\chi_{\rm l})\,\Sigma_{\rm cr}\;,
\end{equation}
so that the comoving critical surface density is smaller by a factor
$a^2(\chi_{\rm l})$. This makes sense: for a given lens, the comoving
surface mass density (mass per unit comoving area) is smaller than the
proper surface mass density, 
\begin{equation}
\Sigma_{\rm com}(R_{\rm com})=a^2(\chi_{\rm l})\,\Sigma(R_{\rm
  com})\;,
\end{equation}
since the comoving area is larger than
the proper one by a factor $a^{-2}$. Correspondingly, since the
convergence, or the correlation function in equation (\ref{eq:4}), is independent
of whether proper or comoving measures are 
used, the comoving critical surface mass density is smaller by the
same factor.

If $N$ lensing galaxies at redshift $z_{\rm l}$ are located at
positions $\boldsymbol{\theta}_i$ within a solid angle $\omega$, the
corresponding fractional number density contrast reads
\begin{equation}
\kappa_{\rm g}(\boldsymbol{\theta})={1\over \overline{n}}\sum_{i=1}^N\delta_{\rm
  D}(\boldsymbol{\theta}-\boldsymbol{\theta}_i) - 1\;,
\label{eq:11}
\end{equation}
where for large $N$ and $\omega$, $\overline{n}=N/\omega$. To evaluate the
correlator of equation (\ref{eq:3}) in this case, we replace the ensemble
average with an angular average, as is necessarily done in any
practical estimation,
\begin{equation}
\ave{\kappa_{\rm g}(\boldsymbol{\theta})\,\kappa(\boldsymbol{\theta}+\boldsymbol{\vartheta})}
\approx 
{1\over \omega}\int_\omega \mathrm{d}^2\theta\;
\kappa_{\rm g}(\boldsymbol{\theta})\,\kappa(\boldsymbol{\theta}+\boldsymbol{\vartheta})
=
{1\over \omega}\int_\omega \mathrm{d}^2\theta\;
\eck{{1\over \overline{n}} \sum_{i=1}^N\delta_{\rm
  D}(\boldsymbol{\theta}-\boldsymbol{\theta}_i)}\,\kappa(\boldsymbol{\theta}+\boldsymbol{\vartheta})
={1\over N}\sum_{i=1}^N\kappa(\boldsymbol{\theta}_i+\boldsymbol{\vartheta})\;,
\label{eq:12}
\end{equation}
valid for separations $\vartheta$ which are much smaller than the
linear angular extent $\sqrt{\omega}$ of the region (to neglect
boundary effects), and we employed
the fact that the ensemble average -- and in the same approximation as
above, the angular average -- of $\kappa(\boldsymbol{\vartheta})$ vanishes. 
We thus see that
the correlator $\ave{\kappa_{\rm g}\kappa}$ can be obtained from the
average convergence around the foreground galaxies, a quantity probed
by the shear. Thus we find the relations
\begin{equation}
\gamma_{\rm t}(\theta)=\Sigma_{\rm cr}^{-1}\,\Delta\Sigma(\chi_{\rm l} \theta)
=\Sigma_{\rm cr,com}^{-1}\,\Delta\Sigma_{\rm com}(\chi_{\rm l} \theta) \;,
\label{13}
\end{equation}
where 
\begin{equation}
\Delta\Sigma(R_{\rm com})={2\over R_{\rm com}^2}\int_0^{R_{\rm com}}
\mathrm{d} R\;R\,\Sigma(R) - \Sigma(R_{\rm com})   \;,
\label{eq:14}
\end{equation}
and the analogous definition for $\Delta\Sigma_{\rm com}$.

A further subtlety and potential source of confusion is that
frequently, the surface mass density $\Sigma$ is considered a
function of {\it proper} transverse separation $R=a(\chi_{\rm l})
R_{\rm com}$. For the purpose of this appendix, we call this function
$\Sigma_{\rm p}$, which is related to $\Sigma$ by
\begin{equation}
\Sigma_{\rm p}(R)=\Sigma[R/a(\chi_{\rm l})]\;, \quad  \hbox{or}
\qquad
\Sigma(R_{\rm com})=\Sigma_{\rm p}[a(\chi_{\rm l})R_{\rm com}]\;,
\label{eq:15}
\end{equation}
yielding
\begin{equation}
\gamma_{\rm t}(\theta)=\Sigma_{\rm cr}^{-1}\,\Delta\Sigma(\chi_{\rm
  l} \theta)
=\Sigma_{\rm cr}^{-1}\,\Delta\Sigma_{\rm p}(D_{\rm l} \theta) \;.
\label{eq:16}
\end{equation}
We argue that the definition used should depend on the science case.
For example, when considering the mean density profile of galaxies, it
is more reasonable to use proper transverse separations -- as that
density profile is expected to be approximately stationary in proper
coordinates. For larger-scale correlations between galaxies and
matter, however, the use of comoving transverse separations is more meaningful,
since the shape of the cross-correlation function on large scales is
expected to be approximately preserved. 

\subsection{Relation to previous work}
In the literature on galaxy-galaxy lensing, one finds relations that
differ from the ones derived above; we shall comment on some of these
differences here.

The first aspect is that in several papers \citep[e.g.][]{Mandelbaum2010, Viola2015, Torre2017}, the
integrand in equation (\ref{eq:4}) is replaced by $1+\xi_{\rm gm}$,
implying that the corresponding $\Sigma_{\rm com}$ contains the
line-of-sight integrated mean density of the Universe, in addition to
the correlated density. This constant term is, however, not justified by
the derivation in Appendix \ref{sec:appendix2deriv}. 
While such a constant drops out in the
definition of $\Delta\Sigma_{\rm com}$, and thus does not impact on
quantitative results, it nevertheless causes a principal flaw: its
inclusion would imply that the convergence $\kappa= \Sigma_{\rm cr,
  com}^{-1} \Sigma_{\rm com}$ for all lines-of-sight to redshifts
$z_{\rm s}\sim 1$ would be several tenths, causing a large difference
between shear and reduced shear [$g = \gamma / (1 + \kappa)$], which is the observable in weak
lensing. That is, it would strongly 
modify the relation between $\gamma$ and the observable image
ellipticities, yielding significant biases in all weak lensing
studies. Indeed, the mean density of the Universe is
already taken into account by the Robertson--Walker metric: the fact
that the angular-diameter distance is a non-monotonic function of
redshift can be considered as being due to the gravitational light
deflection by the mean mass density of the Universe -- the convergence
part in the optical tidal equation \citep[see, e.g.][]{Seitz1994}. 
Howewer, this is usually not called `lensing', but `curvature
of the metric'. Lensing is usually ascribed solely to the effect
caused by density inhomogeneities. But in any case: the mean cosmic
density can not be accounted for twice, once for the metric [and thus
the use of (comoving) angular-diameter distances in a FRW model], and
a second time for the convergence on such a background model.

A second issue in some of the recent GGL papers is mixing the use of
the comoving surface mass density, $\Sigma_{\rm com}$ (equation \ref{eq:7}), with the proper
critical surface mass density, $\Sigma_{\rm cr}$ (equation \ref{eq:9}). For
example, \citet{Torre2017} in their equation (18) (apart from
the constant term discussed above) define $\Sigma_{\rm com}$ as in equation (\ref{eq:9}), but use in their equations (9) and (10) the proper
critical surface mass density $\Sigma_{\rm cr}$ to relate the
tangential shear to $\Sigma_{\rm com}$. Hence, this relation would cause an
offset by a factor $(1+z_{\rm l})^2$ from the correct result. However, the inconsistency appears only in the text of the paper and not in the code or calculations performed \citep[][private communication]{Torre2018}.

The same issue occurs in the write-up in several earlier publications
of our KiDS team. For example, the equations (2,5,6) in \citet{Viola2015} show this inconsistency (where we also point out a
typo in the integration limits of equation 2), as well as equations
(1,2,3) in \citet{VanUitert2016} and, as mentioned already in the
main text, equations (1,2,3) in \citet{Dvornik2017a}. We have
checked the codes that were used to derive the quantitative results in
these papers to see whether they employ the same inconsistent use of
quantities. We found that there is an inconsistency present only in writing, namely in equation (2) of \citet{Viola2015} and not in the code that was used to produce the results. As pointed out above the correlation function is given in comoving coordinates, while the extraction of the galaxy-galaxy lensing signal is calculated in proper coordinates. Thus, the equation (2) of \citet{Viola2015} should correctly read as (using their notation):
\begin{equation}\label{eq:correct_viola}
\Sigma(R) = 2 \overline{\rho}_{\text{m}} \left(1 + \langle z_{\text{l}} \rangle \right)^{2} \int_{0}^{\pi_{\text{s}}} \xi_{\text{gm}} (\sqrt{R^{2} + \Pi^{2}}) \mathrm{d} \Pi \,.
\end{equation}
Put differently, while the data were indeed extracted in proper coordinates, the outputs of the halo model were not reported to be scaled to the coordinates used by the data \citep[][private communication]{Cacciato2017}. Secondly, the equations (6) and (9) in \citet{Dvornik2017a} should have the same form as equation (\ref{eq:SigmaCrit}) and (\ref{eq:crit_effective}) in this paper, again having a mistake only in writing. The exact same correction should also be applied to the equation (3) of \citet{Velliscig2016}.

The erroneous formulation also occurred in a previous paper from \citet{VanUitert2016}. Their equation (2), should read correctly as equation (\ref{eq:SigmaCrit}), which is (again, using their notation):
\begin{equation}
\label{eq:SigmaCrit2}
\Sigma_{\text{crit}}=\frac{c^2}{4\pi G (1+ z)^{2}} \frac{D_{\rm S}}{D_{\rm L}D_{\rm LS}} \, .
\end{equation}
Further KiDS analyses from \citet{Amon2017} and \citet{Amon2017a} measure large-scale galaxy-mass correlations using the comoving critical surface mass density. The definitions, presented in these two papers are consistent with the data and the equations derived in Appendix \ref{sec:appendix2deriv}.

On the other hand, the KiDS analyses from \citet{Sifon2015}, \citet{Brouwer2016} and \citet{Brouwer2016a} use a separate NFW stacking method that does not rely on the halo model, and all use proper coordinates that are consistent with the data.


\bsp	
\label{lastpage}
\end{document}